\documentclass[twocolumn,prd,preprintnumbers,amsmath,amssymb,nofootinbib,superscriptaddress]{revtex4-1}
\usepackage{mathrsfs}
\usepackage[dvipsnames]{xcolor}
\newcommand{\lagrange}[1]{{\color{blue} $#1$ } }
\newcommand{\Lie}[1]{\ensuremath{{\cal L}_{#1}}}

\newcommand{\spartial}{\ensuremath{\hat{\partial}}}
\newcommand{\grad}{\ensuremath{\vec{\nabla}}}

\newcommand{\mub}{{\bar{\mu}}}
\newcommand{\nub}{{\bar{\nu}}}
\newcommand{\phib}{{\bar{\phi}}}

\newcommand{\neta}{{\delta \nu}}
\newcommand{\varpit}{\tilde{\varpi}}
\newcommand{\netat}{{ (\delta \nu)^2}}
\newcommand{\lambdas}{\lambda_s}
\newcommand{\eps}{\epsilon}

\newcommand{\alphat}{{\tilde{\alpha}}}
\newcommand{\Rcal}{{\cal R}}
\newcommand{\Kcal}{{\cal K}}

\newcommand{\coordT}[1]{\ensuremath{{\vec{#1}}}}

\newcommand{\Fh}{\hat{F}}

\newcommand{\Ah}{\hat{A}}
\newcommand{\Pib}{\bar{\Pi}}

\newcommand{\Hpri}{H_{\rm pri}}
\newcommand{\HcalP}{{\cal H}_{\rm pri}}
\newcommand{\Hsec}{H_{\rm sec}}
\newcommand{\Htot}{H_{\rm FC}}

\newcommand{\PA}{\Pi^{(A)}}

\newcommand{\Pmu}{\Pi^{(\mu)}}
\newcommand{\Pnu}{\Pi^{(\nu)}}
\newcommand{\lambdamu}{\lambda^{(\mu)}}
\newcommand{\lambdanu}{\lambda^{(\nu)}}
\newcommand{\lambdaA}{\lambda^{(A)}}
\newcommand{\lambdaB}{\lambda^{(B)}}
\newcommand{\Poisson}[2]{\left\{#1,#2\right\}}

\newcommand{\smear}[2]{{#1}[{#2}]}
\newcommand{\Constr}{{\cal C}}

\newcommand{\Hcal}{{\cal H}}

\newcommand{\Ucal}{{\cal U}}
\newcommand{\Fcal}{{\cal F}}

\newcommand{\Qcalb}{\bar{{\cal Q}}}
\newcommand{\Qcal}{{\cal Q}}
\newcommand{\Ycal}{{\cal Y}}

\newcommand{\Acal}{{\cal A}}
\newcommand{\Bcal}{{\cal B}}

\newcommand{\Jcal}{{\cal J}}

\newcommand{\Scal}{{\cal S}}
\newcommand{\Smu}{\Scal^{(\mu)}}
\newcommand{\Snu}{\Scal^{(\nu)}}
\newcommand{\SA}{\Scal^{(A)}}
\newcommand{\SB}{\Scal^{(B)}}
\newcommand{\umu}{u^{(\mu)}}
\newcommand{\unu}{u^{(\nu)}}
\newcommand{\uA}{u^{(A)}}
\newcommand{\uB}{u^{(B)}}

\newcommand{\xiA}{\xi^{(A)}}
\newcommand{\xiB}{\xi^{(B)}}
\newcommand{\qh}{\hat{q}}

\newcommand{\Gt}{\ensuremath{\tilde{G}}}
\newcommand{\GN}{\ensuremath{G_{{ N}}}}
\newcommand{\KB}{\ensuremath{K_{{ B}}}}

\newcommand{\ph}{\phantom}
\newcommand{\nn}{\nonumber}

\begin{document}

\title{Aether scalar tensor theory: Hamiltonian Formalism}

\author{Marianthi Bataki}
\email{bataki.marianthi@ucy.ac.cy}
\affiliation{Department of Physics, University of Cyprus
1, Panepistimiou Street, 2109, Aglantzia, Cyprus}                           
\affiliation{ CEICO, Institute of Physics of the Czech Academy of Sciences, Na Slovance 1999/2, 182 00, Prague, Czechia }             
\author{Constantinos Skordis}
\email{skordis@fzu.cz}
\affiliation{ CEICO, Institute of Physics of the Czech Academy of Sciences, Na Slovance 1999/2, 182 00, Prague, Czechia }             
\affiliation{Department of Physics, University of Oxford, Denys Wilkinson Building, Keble Road,   Oxford OX1 3RH, UK}
\author{Tom Zlosnik}
\email{thomas.zlosnik@ug.edu.pl}
\affiliation{ Institute of Theoretical Physics and Astrophysics, University of Gdańsk, 80-308 Gdańsk, Poland }
\date{\today}

\begin{abstract}                                              
The Aether Scalar Tensor (AeST) theory is an extension of General Relativity (GR), proposed for addressing galactic and cosmological observations without dark matter.
 By casting the AeST theory into a $3+1$ form, we determine its full non-perturbative Hamiltonian formulation and analyse the resulting constraints. 
We find the presence of four first class and four second class constraints and show that the theory has six physical degrees of freedom at the fully nonlinear level. 
Our results set the basis for determining the propagation of perturbations on general backgrounds and we present the case 
of small perturbations around Minkowski spacetime as an example stemming from our analysis.
\end{abstract}

\maketitle

\section{Introduction}
\label{sec:introduction}
Einstein's theory of General Relativity (GR) has been extraordinarily successful and remains the paradigmatic theory of gravity.
 Data from a wide range of astrophysical systems, from precision tests of gravity in the solar system \cite{Will2014} to recent measurements of 
gravitational waves \cite{TheLIGOScientific:2017qsa} are in accord with its predictions. It is expected that 
corrections to GR will become important at extremely high energy scales/short length scales (for instance in the early universe \cite{Starobinsky:1980te}),
however, a question mark hangs over whether additional structure in the gravitational sector may play a prominent role on cosmological and certain astrophysical scales. 

Extending GR with additional structure relevant in the low gradient/curvature regime --the conditions on galactic and cosmological scales-- is an 
 intriguing possibility considering that, otherwise, immense observational evidence points to additional non-baryonic matter driving the gravitational dynamics in those regimes: dark matter.
 A notable early example are the observations that rotation curves of spiral galaxies are asymptotically flat~\cite{RubinEtAl1980,Persic:1995ru}, 
captured more recently by the radial acceleration 
relation (RAR)~\cite{McGaughEtAl2016,Lelli:2016cui}. Assuming GR, this is only possible if galaxies are immersed within dark matter halos. Dark matter halos are
seen to be even more prominent concerning dwarf and ultra-faint dwarf galaxies~\cite{Simon:2019nxf}. On larger scales, dark matter is
necessary in explaining observations of galaxy clusters~\cite{Hahn:2006mk}, 
weak lensing tomography~\cite{Hoekstra:2008db}, cluster lensing~\cite{Natarajan:2024iqm} and galaxy-galaxy~\cite{Yoo:2005eh} strong lensing. 
Notable are the cases of merging galaxy clusters, indicating an offset of the baryonic mass seen through x-rays,
and the dynamical mass seen through lensing~\cite{Clowe_2006}. Finally, at the largest scales, the observed clustering of galaxies and voids, e.g.~\cite{eBOSS:2017dtq,eBOSS:2021pff},
and the cosmic microwave background~\cite{Planck:2018vyg} indicate five times more dark matter than baryonic matter.
The $\Lambda$-cold dark matter ($\Lambda$CDM), where dark matter is modelled as a distribution of collisionless particles with cold initial conditions, is the simplest model which 
fits the totality of the data (although a few tensions with $\Lambda$CDM have emerged in the recent years, e.g.~\cite{Knox:2019rjx,Amon:2022azi,Sakr:2021jya,Secrest:2022uvx}).

Despite there being many proposed candidates for what dark matter may be, see \cite{Jungman:1995df,Bertone:2004pz,Marsh:2015xka,Green:2020jor} for reviews, 
and several experiments searching for particle dark matter either 
through nuclear recoil, see~\cite{LZ:2022lsv,XENON:2023sxq} or astrophysical production mechanisms~\cite{AMS:2019rhg,Planck:2018vyg}, 
the actual particle is currently undetected. 
Thus, there remains the possibility that what is being observed 
may not be the effect of the presence of  dark matter, but that of additional gravitational degrees of freedom leading to
a change of the way known matter affects the gravitational field. For this to manifest, an extension of GR must be at play.

Modified Newtonian Dynamics (MOND) is a non-relativistic framework proposed by Milgrom~\cite{Milgrom1983a}, as a way of addressing galactic observations without dark matter.
In one formulation, Newton's second law of motion is changed below an acceleration scale $a_0\sim 1.2 \times 10^{-10} m/s^2$ while non-relativistic gravity is governed by Poisson's equation.
In another formulation, Newton's second law is kept but the gravitational equation determining the non-relativistic potential $\Phi$ from the matter density is generalized
and departs from Poisson's equation at low potential gradients determined by $a_0$. 
Several instances of the second formulation exist, starting from a single potential formulation of Bekenstein and Milgrom~\cite{BekensteinMilgrom1984}, to
other later formulations involving additional potentials~\cite{Milgrom:2009ee,Milgrom:2023idw}.
The RAR~\cite{McGaughEtAl2016,Lelli:2016cui} and the (baryonic) Tully-Fisher relation~\cite{Tully:1977fu,McGaugh:2000sr} 
comfortably emerge within the MOND paradigm, lending additional support for investigating this possibility furher.

The MOND proposal leads to a wide variety of predictions that can be compared against astrophysical data \cite{FamaeyMcGaugh2011,Banik:2021woo} though has the inherent
 restriction that in the absence of a relativistic completion, it has not been clear what the realm of validity of MOND is.
This has prompted the construction of a variety of extensions of GR~\cite{BekensteinMilgrom1984,Bekenstein1988,Sanders1997,Bekenstein2004,NavarroVanAcoleyen2005,ZlosnikFerreiraStarkman2006,Sanders2005,Milgrom2009,BabichevDeffayetEsposito-Farese2011,DeffayetEsposito-FareseWoodard2011,BlanchetMarsat2011,Sanders2011,Mendoza:2012hu,Woodard2014,Khoury2014,Hossenfelder2017,Burrage:2018zuj,Milgrom:2019rtd,DAmbrosio:2020nev,Blanchet:2023vln,Kading:2023hdb},
which lead to  MOND behaviour at the quasistatic weak-field limit\footnote{
Extended dark matter models have also been proposed for accommodating some of the MOND phenomenology, and 
which have $\Lambda$CDM behaviour on a Friedman-Lemaitre-Robertson-Walker (FLRW) Universe plus linear fluctuations. We enumerate some of these here.
The dipolar dark matter model~\cite{Blanchet:2006yt,Blanchet:2009zu} leads to MOND behaviour in galaxies, while predicting
 novel behaviour such as time-varying non-Gaussianities~\cite{Blanchet:2012ey}, but has been shown to have an instability which may lead to
the evaporation of galaxies~\cite{Stahl:2022vaw}. In~\cite{Berezhiani:2015pia,Berezhiani:2015bqa} dark matter has a superfluid phase whose exhitations (phonons) lead to MOND in galaxies 
while retaining $\Lambda$CDM behaviour cosmologically. The self-interacting dark matter model~\cite{Kaplinghat:2015aga,Kamada:2016euw} has been shown to accommodate the RAR and predicts 
cored halo profiles in contrast with pure CDM haloes which in $\Lambda$CDM have cuspy profiles~\cite{Navarro:1995iw}. Another class of models are based on a dark matter-baryon 
interaction~\cite{Famaey:2017xou}, also retaining $\Lambda$CDM behaviour on the largest scales while recovering the RAR and other MOND phenomenology in galaxies.
}.
 
However, none of these extensions have been shown to fit the observations of the Cosmic Microwave Background radiation as most recently reported by the Planck Surveyor satellite~\cite{Planck:2018vyg,Tristram:2023haj}. 
Moreover - although exceptions may be found - they usually do not lead to a gravitational wave tensor mode speed equalling the speed of light as required by data~\cite{TheLIGOScientific:2017qsa,Savchenko:2017ffs}.

Starting from a general class of theories based on a metric, a unit-timelike vector field and a scalar field~\cite{SkordisZlosnik2019},
the AeST theory~\cite{SkordisZlosnik2020} was constructed 
to have a massless spin-2 graviton which propagates at the speed of light while tending to MOND in the weak-field quasistatic regime relevant to galaxies.
The property of the vector field to be unit-timelike is important for employing the Sanders mechanism~\cite{Sanders1997} to lead to correct gravitational lensing for isolated masses in the absense 
of dark matter~\cite{SkordisZlosnik2020}.
It was further required to have a Friedman-Lemaitre-Robertson-Walker (FLRW) behaviour extremely close to that of the $\Lambda$CDM model by
having features akin to shift-symmetric k-essense~\cite{Scherrer2004} and ghost condensate theory~\cite{ArkaniHamedEtAl2003,ArkaniHamed:2005gu}.
The closeness to $\Lambda$CDM persists also when linear fluctuations around FLRW are included, 
which enable the theory to provide a similarly good match to precision cosmological data (for example the linear matter power spectrum and CMB temperature and polarization anisotropy power spectra)
 to the $\Lambda$CDM model.
It was further shown that linear fluctuations on  Minkowski spacetime
propagate two massless tensor modes, two massive vector modes and one massive scalar mode, all of which are healthy provided certain
constraints on the theory parameters are satisfied~\cite{SkordisZlosnik2021}. A sixth mode was shown to have a linear $t$-dependence and to have positive Hamiltonian for
momenta larger than a mass scale which observationally is $\lesssim 10^{-30} eV$, and negative otherwise. This  behaviour is akin to a Jeans instability
and does not cause quantum vacuum instability at low momenta~\cite{GumrukcuogluMukohyamaSotiriou2016}.  Further studies of the AeST theory 
have been performed in ~\cite{Bernardo:2022acn,Kashfi:2022dyb,Mistele:2021qvz,Mistele:2023paq,Tian:2023gjt,Llinares:2023lky,Verwayen:2023sds,Mistele:2023fwd}.
We also note that the new Khronon proposal of~\cite{Blanchet:2024mvy} shares several features of AeST theory and can also fit the large scale cosmology, however, it is simpler in that it does not contain a vector field.

Despite these promising features of the theory, it is crucial that it can match the success of GR in all cases where it has been tested, while fitting observations in the regimes where a successful account of the data in the context of GR and known matter requires the addition of dark matter.  This will require finding solutions to the theory in systems that might not be describable by linear perturbations propagating on highly symmetric backgrounds. 
Towards these ends, an important first step will be to cast the equations as first order evolution equations in time  - this will enable both analytical and numerical solutions for more complicated situations
to be more easily found.

In this paper we develop the canonical/Hamiltonian formulation of the theory, following the Dirac-Bergman 
formulation~\cite{Bergmann:1949zz,Anderson:1951ta,Dirac:1950pj,Dirac:1958sc,dirac2001lectures,Henneaux:1992ig} 
which was developed in the case of GR~\footnote{
The Hamiltonian formulation of other theories beyond GR has been studied elsewhere, such as, $D=10$ supergravity~\cite{Henneaux:1986cz}, the Plebanski theory~\cite{Buffenoir:2004vx},  $f(R)$ theories~\cite{Deruelle:2009zk}, 
the Tensor-Vector-Scalar (TeVeS) theory ~\cite{Chaichian:2014dfa}, the Degenerate Higher-Order Scalar-Tensor theories (DHOST) which include Horndeski and beyond-Horndeski theories~\cite{Lin:2014jga,Langlois:2015skt},
and the minimal varying $\Lambda$ theories~\cite{Alexandrov:2021qry}.
The Hamiltonian formulation has also been used to study the Bondi-Metzner-Sachs group at spatial infinity in the case of GR~\cite{Henneaux:2018cst}.
}.
This allows us to put the theory's equations of motion in the form of Hamilton's first-order equations of motion. 
The completion of the canonical analysis also enables clarification of other issues, such as, the number of degrees of freedom that the theory possesses and whether the theory is an example 
of an irregular system, that is, a theory where the canonical structure varies throughout phase space.
A manifestation of the latter can be that perturbations around some backgrounds describe different number of degrees of freedom than perturbations around 
other backgrounds, see~\cite{Alexandrov:2012yv,Alexandrov:2021qry} for examples.

The outline of the paper is as follows: In Section \ref{The_Theory} we introduce the theory; 
in Section \ref{ADM_formalism} we introduce the Arnowitt-Deser-Misner (ADM)~\cite{Arnowitt1962} formalism and apply it to the AeST theory and its associated decomposition of fields; 
in Section \ref{Hamiltonian_analysis} we cast the theory in Hamiltonian form and perform a full constraint analysis in section \ref{Constraint_analysis}; 
in Section \ref{Minkowski_example} we restrict the full non-perturbative Hamiltonian to the case of small perturbations 
around a Minkowski spacetime solution and in doing so demonstrate the recovery of the results previously found in~\cite{SkordisZlosnik2021}.
Finally, in Section \ref{Conclusions} we present our conclusions.

\section{The theory}
\label{The_Theory}

The theory depends on a metric $g_{\mu\nu}$ universally coupled to matter so that the Einstein equivalence principle is obeyed, a scalar field $\phi$ and 
a unit time-like vector field $\Ah^\mu$~\footnote{We depart from previous expositions of this theory~\cite{SkordisZlosnik2020,SkordisZlosnik2021} and denote the 4-dimensional vector field by $\Ah^\mu$. We reserve the
symbol $A^\mu$ (without the `hat') for the projected vector field on the 3-dimensional hypersurface  introduced in section \ref{ADM_formalism}, see equation \eqref{A_mu_decomposition},
 as this will feature more prominently  than $\Ah^\mu$ in the present work.}, where the unit-timelike condition is enforced by a Lagrange multiplier $\lambda$. The action is 
\begin{widetext}
\begin{align}
S 
=& 
 \int d^4x \frac{\sqrt{-g}}{16\pi \Gt}\bigg\{ R  - 2 \Lambda
 - \frac{\KB}{2}  \Fh^{\mu\nu} \Fh_{\mu\nu} 
+   (2-\KB) \left( 2J^{\mu} \nabla_\mu \phi - \Ycal\right)
- \Fcal(\Ycal,\Qcal)
 - \lambda(\Ah^\mu \Ah_\mu+1)
\bigg\} 
 + S_m[g]
\label{NT_A_action}
\end{align}
\end{widetext}
where $g$ is the metric determinant, $\nabla_\mu$ the covariant derivative compatible with $g_{\mu\nu}$,   $R$  is the Ricci scalar, $\Lambda$ is the cosmological constant,
$\Gt$ is the bare gravitational strength, $\KB$ is a constant and $\lambda$ is a Lagrange multiplier imposing the unit time-like constraint on $A_{\mu}$. 
We adopt the $(-,+,+,+)$ metric signature convention and - unless otherwise specified - we employ the Einstein summation convention.
In addition, we have defined the tensors $J^\mu = \Ah^\nu \nabla_\nu \Ah^\mu$ and $\Fh_{\mu\nu} = 2\nabla_{[\mu} \Ah_{\nu]}$ 
while the matter action $S_m$ is  assumed not to depend explicitly on $\phi$ or $\Ah^\mu$.
Defining 
\begin{align}
\qh_{\mu\nu}  \equiv & g_{\mu\nu}  + \Ah_\mu \Ah_\nu
\label{q_hat}
\end{align}
the function $\Fcal(\Ycal,\Qcal)$ depends on the scalars 
\begin{align}
\Qcal \equiv & \Ah^\mu \nabla_\mu \phi
\label{Q_cal}
\end{align}
and
\begin{align}
\Ycal \equiv& \qh^{\mu\nu} \nabla_\mu \phi \nabla_\nu \phi
\label{Y_cal}
\end{align}
The function $\Fcal$ is subject to conditions so that the  cosmology of the theory is compatible with $\Lambda$CDM on
FRLW spacetimes and a MOND limit emerges in quasistatic situations~\cite{SkordisZlosnik2020}.
The choice of the action (\ref{NT_A_action}) is largely informed by the phenomenological requirements discussed in Section \ref{sec:introduction}, starting from
a more general action (19) of \cite{SkordisZlosnik2019} which depends on eight free functions $d_i$. As discussed in~\cite{SkordisZlosnik2019}, setting $d_1+ d_3 =0$
is sufficient to ensure a tensor mode propagating at the speed of light on any background. Several other constraints among the $d_i$s ensure MOND behaviour on a quasistatic background as in TeVeS theory 
and FLRW cosmology with the AeST fields behaving as dust.

On a flat FLRW background the metric takes the form $ds^2 = -dt^2 + a^2 \gamma_{ij} dx^i dx^j$ where $a(t)$ is the scale factor and $\gamma_{ij}$ is a flat spatial metric.
The vector field reduces to $\Ah^\mu = ( 1, 0, 0, 0)$ while $\phi \rightarrow \phib(t)$ leading to $\Qcal\rightarrow \Qcalb = \dot{\phib}$ and  $\Ycal \rightarrow 0$,
so that we may define $\Kcal(\Qcalb) \equiv -\frac{1}{2} \Fcal(0,\Qcalb)$.
We require that $\Kcal(\Qcalb)$ has a minimum at $\Qcal_0$ (a constant) so that we may expand it as $\Kcal = \Kcal_2\left(\Qcalb - \Qcal_0\right)^2 + \ldots$, where the $(\ldots)$ denote 
higher terms in this Taylor expansion.
This condition leads to $\phib$ contributing energy density scaling as dust $\sim a^{-3}$ akin to~\cite{Scherrer2004,ArkaniHamedEtAl2003},  plus small corrections which tend 
to zero when $a\rightarrow \infty$. In principle, $\Kcal$ could be offsetted from zero at the minimum $\Qcal_0$, i.e. $\Kcal(\Qcal_0) = \Kcal_0$,
 however, such an offset can always be absorbed into the cosmological constant $\Lambda$ and thus we choose $\Kcal_0 = 0$ by convention, implying the same on the parent function $\Fcal$.

In the quasistatic weak-field limit we may set the scalar time derivative to be at the minimum $\Qcal_0$, as is expected to be the case in the late universe.
This means that we may expand $\phi = \Qcal_0 t + \varphi$. Moreover, in this limit $\Fcal \rightarrow  (2-\KB) \Jcal(\Ycal)$, with $\Jcal$ defined appropriately as
$\Jcal(\Ycal) \equiv \frac{1}{2 -\KB} \Fcal(\Ycal,\Qcal_0)$.
It turns out that MOND behaviour emerges if $\Jcal \rightarrow \frac{2\lambdas}{3(1+\lambdas) a_0} |\Ycal|^{3/2}$ where
$a_0$ is Milgrom's constant and $\lambdas$ is a constant which is related to the Newtonian/GR limit. Specifically, there are two ways that GR can be restored: (i) screening 
and (ii) tracking.
In the former, the scalar is screened at large gradients $\grad_{i} \varphi$, where $\grad_i$ is the spatial gradient on a flat background space with metric $\gamma_{ij}$,
and in the latter, $\lambdas \varphi$ becomes proportional to the Newtonian potential, leading to an effective Newtonian constant 
\begin{equation}
\GN = \frac{1 + \frac{1}{\lambdas}}{ 1 - \frac{\KB}{2} } \Gt.
\end{equation}
Screening may be achieved either through terms in $\Jcal \sim \Ycal^p$ with $p>3/2$ or through galileon-type terms which must be added to \eqref{NT_A_action}. Either way, for our purposes in this article,
we may model screening as $\lambdas \rightarrow \infty$. 

We conclude this section by comparing the AeST model to other extensions of GR.
It is straightforward to extend GR with the addition of a scalar field  and several scalar-tensor theories have been
proposed to play a role of Dark Energy (DE).  While several models exist, they generally fall under the general Degenerate Higher-Order Scalar-Tensor 
theories (DHOST)~\cite{Langlois:2015skt}, which include Horndeski~\cite{Horndeski:1974wa,Deffayet:2009mn}  and beyond-Horndeski theories~\cite{Gleyzes:2014dya}.
These are not Effective Field Theories (EFT) in the strict sense but are covariant theories leading to at most 2nd order field equations.
The Effective Field Theory of DE (EFTofDE) (see~\cite{Frusciante_2020} for a review) is constructed on a general FLRW background plus linearized perturbations by including
all possible terms at that order that may arise from the metric perturbation, or a scalar field (typically written in the unitary gauge).
 The  majority of the terms which are part of DHOST, or EFTofDE, do not overlap with the AeST action \eqref{NT_A_action}. 
The only term that may overlap with DHOST is $\Fcal(\Ycal,\Qcal) + (2-\KB) \Ycal $ and only in specific case where the vector field can be ignored (e.g. FLRW cosmology).
 Nevertheless, AeST could in principle be extended with additional terms for $\phi$ coming from DHOST and obeying the shift symmetry.

Extending GR with a vector field is another direction that has been considered. In~\cite{Heisenberg:2014rta} GR was extended with a massive vector field which generalizes the Proca action
(yet another generalization of Proca theory was studied more recently in~\cite{deRham:2020yet} whilst novel couplings between a Proca field and dark matter were considered in detail in \cite{Gomez:2022okq} ). In~\cite{Heisenberg:2018acv}, a Scalar-Vector-Tensor (SVT) theory  was proposed which
blends together Horndeski and generalized Proca theories; see~\cite{Heisenberg_2019} which reviews DHOST and its subsets, generalized Proca, SVT and other theories.
In all those theories, the vector field is not necessarily unit-timelike as required by AeST and so there is almost no obvious overlap with AeST (apart from the case described above, related to DHOST).
We note, however, that the SVT with broken gauge-invariance has six propagating degrees of freedom which is the same as in AeST theory, as we show below. Thus it would be interesting to further
probe a possible connection between the two, although there is no guarantee that there is a concrete connection.

Theories which are mostly related to AeST are those of ghost condensate~\cite{ArkaniHamedEtAl2003,ArkaniHamed:2005gu} and gauge ghost condensate~\cite{Cheng:2006us}, also called bumblebee model 
in \cite{Kostelecky:1988zi,Kostelecky:1989jw}. The vector field in~\cite{Cheng:2006us,Kostelecky:1988zi,Kostelecky:1989jw} is also not unit-timelike, however, it has a symmetry-breaking potential
which spontaneously breaks time-diffeomorphisms at its minimum. Indeed, it can be shown~\cite{Cheng:2006us} that in the decouling limit the gauge condensate theory becomes the Einstein-Aether 
theory~\cite{JacobsonMattingly2001}, discovered earlier by Dirac~\cite{Dirac1962}, which lends to AeST the $F_{\mu\nu} F^{\mu\nu} + \lambda (A_\mu A^\mu+1)$ term. 

Rather than extending GR, several models have been proposed for studying possible extensions of CDM by using a parametrized approach to encompass as large a landscape of models as possible
 rather than specifically referring to particular theories.  The effective theory of structure formation (ETHOS)~\cite{Cyr-Racine:2015ihg} is a model that encompasses general interactions
 of dark matter with a dark radiation component at the FLRW and linearized cosmological regime, plus dark matter self-interactions in the non-linear regime.
 Similarly to ETHOS, the Generalized Dark Matter (GDM) model~\cite{Hu:1998kj,Kopp:2016mhm,Ili__2021} extends CDM in the linearized regime by
letting dark matter have a general time-dependent equation of state, sound speed and viscosity~\footnote{GDM has some overlap with 
the linearized regime of ETHOS as the former can emerge by treating tightly-coupled fluids as a single fluid~\cite{Kopp:2016mhm}, amongst other possibilities.}.  
There is no unique non-linear completion to GDM and specific theories with GDM limit include ultra-light axions 
and the Khronon theory~\cite{Blanchet:2024mvy}  (which is a GR  rather than a CDM extension). 
Neither is ETHOS nor is GDM an EFT theory in the usual sense.
The Effective Field Theory of Large Scale Structures (EFTofLSS), is however, a rigorous classical EFT in the usual sense~\cite{BaumannNicolisSenatoreEtal2012,Carrasco:2012cv,Porto:2013qua}, 
particularly suited for parameterizing the mildly non-linear regime of CDM, and can be extended to include other theories beyond CDM, see e.g.~\cite{Cusin:2017wjg,Carrilho:2022mon}.
Neither MOND, nor AeST is captured by the above formalisms (however, the Khronon theory~\cite{Blanchet:2024mvy} does fall under GDM cosmologically).

\section{3+1 formalism}
\label{ADM_formalism}
\subsection{ADM decomposition}
 
\subsubsection{Decomposition of the metric and its derivatives}
As a necessary first step towards constructing the Hamiltonian formalism for the theory, we must make a distinction between space and time. Specifically, we follow the ADM~\cite{Arnowitt1962} formalism
and assume that for the region of spacetime of interest, there exists a global time coordinate $t(x^{\mu})$.
Given this, we may define a `flow of time' vector field $t^{\mu}$ which satisfies $t^{\mu}\nabla_{\mu}t=1$. We  use the notation $\dot{f} \equiv \partial_{t}f$ for some field $f$. Furthermore, we may define a vector field $n^{\mu}$ which is 
normal to surfaces of constant $t$; as such, this field is timelike and may be defined so that it has unit-norm i.e. $g_{\mu\nu}n^{\mu}n^{\nu}=-1$. We may expand this time field
as $t^\mu = N n^\mu + N^\mu$, where we have introduced the lapse function $N$ and shift vector $N^\mu$, which are given respectively by $N = -t^\mu n_\mu$ and $N_\mu = q_{\mu\nu} t^\nu$.
 We coordinatize surfaces of constant $t$ by spatial coordinates $x^{i}$, where $i,j,k$ will be used throughout to denote spatial coordinate indices. The full spacetime metric may be decomposed as:
 
\begin{equation}
g_{\mu\nu} = -n_\mu n_\nu + q_{\mu\nu},
\label{q_proj}
\end{equation}
 where $q_{\mu\nu}$ is the metric on the spatial hypersurface (and therefore, for example $q_{\mu\nu}n^{\mu}=0$).
Note the difference between $\qh_{\mu\nu}$ defined in \eqref{q_hat} and $q_{\mu\nu}$ defined in \eqref{q_proj}.

It is useful to define a spatial derivative $\spartial_{\mu} = q_{\mu}^{\;\;\nu}\partial_{\nu}$ and  covariant derivative $D_\mu$ compatible 
with $q_{\mu\nu}$, i.e. $D_\alpha q_{\mu\nu} =0$. 
Specifically:
\begin{equation}
 D_\mu q_{\alpha\beta} = \spartial_\mu q_{\alpha\beta} - \gamma^{\nu}_{\mu\alpha} g_{\nu\beta} - \gamma^{\nu}_{\mu\beta} g_{\alpha \nu} = 0
\end{equation}
where $\gamma^{\mu}_{\alpha\beta}$ are Levi-Civita symbols associated with the metric $q_{\mu\nu}$ and derivative $\spartial_{\mu}$.  Given $q_{\mu\nu}$ and $n^{\mu}$, a further useful quantity is extrinsic curvature tensor $K_{\mu\nu}$, defined as
\begin{align} 
K_{\mu\nu} \equiv \frac{1}{2} \Lie{n} q_{\mu\nu} = q_\mu^{\;\;\alpha} \nabla_\alpha n_\nu
\end{align}
In component form, we need
\begin{align}
K_{ij} &=\frac{1}{2N} \left(\dot{q}_{ij}-D_iN_j-D_jN_i \right)
\end{align}
while the components of the metric, $n^\mu$ and the Christoffel connection are displayed in appendix \ref{UsefulResults}.

We will adhere to the convention that spatial indices are always lowered and raised with the spatial metric $q_{ij}$, i.e. $K^i_{\;\;j} = q^{ik} K_{kj}$.

\subsubsection{Decomposition of the vector field}
For the vector field $\hat{A}_{\mu}$ we consider a similar decomposition
\begin{equation}
\Ah_\mu =  \chi n_\mu + A_\mu
\label{A_mu_decomposition}
\end{equation}
where $A_\mu \equiv q_{\mu}^{\ph{\mu}\nu} \Ah_\nu$ and 
\begin{equation}
\chi  = -n^\mu \Ah_\mu 
\label{chi_def}
\end{equation}
In component form we find
\begin{align}
\Ah_0 =&  -N \chi  + N^i A_i \;,
 &
\Ah_i =&  A_i
\\
\Ah^0 =&  \frac{\chi}{N} \;,
&
\Ah^i =& A^i -  \frac{\chi}{N} N^i
\end{align}
leading to $ N^i A_i= A_0$, while
\begin{equation}
\chi   = \sqrt{1+|\vec{A}|^2} 
\label{chi_def_comp}
\end{equation}
where $|\vec{A}|^2 \equiv A^i A_i = q^{ij} A_i A_j$, and have taken the positive sign of the square root by convention.

\subsection{The $3+1$ action}
We now present the necessary steps in writing the action \eqref{NT_A_action} in $3+1$ form.
One of the steps involves solving for $\dot{\phi}$ in terms 
of the canonical momenta $\delta S/\delta \dot{\phi}$ and this will involve having to invert potentially very complicated combinations of funtions $\partial \Fcal/\partial \Ycal$
 and $\partial \Fcal/\partial \Qcal$. Instead we can move this structure into elsewhere in the theory by introducing auxiliary fields $\mu$ and $\nu$ such that we set
\begin{align}
\Fcal(\Ycal,\Qcal) = -\nu \Qcal^{2} + \mu \Ycal + \Ucal(\nu,\mu)
\label{Fcal_to_Ucal}
\end{align}
For the scalar field $\phi$, we then find the scalars $\Qcal$ and $\Ycal$  as
\begin{align}
\Qcal &=  \chi \sigma  + A^{i}D_{i}\phi
\label{eq_Qcal}
\\
\Ycal &= |\vec{A}|^2 \sigma^2 + 2\chi \sigma A^i  D_i\phi + \left( q^{ij} +   A^i  A^j \right)  D_i\phi D_j \phi 
\label{eq_Ycal}
\end{align}
where we have defined
\begin{align}
\sigma &= \frac{1}{N}\left( \dot{\phi} - N^{i}D_{i}\phi\right)
\end{align}
Consider now the vector-dependent terms in \eqref{NT_A_action} involving $\Fh_{\mu\nu}$ and $J^\mu$. These depend on the derivatives of $\Ah^\mu$ which are displayed in appendix \ref{UsefulResults}. Using those relations and letting 
\begin{equation} 
F_{ij} \equiv 2 D_{[i} A_{j]} = \Fh_{ij}
\end{equation}
and
\begin{align}
F_i \equiv \frac{1}{N}\Fh_{0i} =& \frac{1}{N} \left[ \dot{A}_i +   D_i \left(N \chi - N^j  A_j\right) \right],
\end{align}
we define the ``magnetic'' aspect of $A^i$ as 
\begin{align}
B^k = \frac{1}{2} \epsilon^{kij} F_{ij},
\end{align}
with inverse $F_{ij} = \epsilon_{ijk} B^k$ and the ``electric'' aspect of $A^i$ as 
\begin{align}
E_i = F_i + \frac{1}{N} \epsilon_{ijk} N^j B^k.
\end{align}
With the above relations and again using \eqref{J_0} and \eqref{J_i}  we find
\begin{align}
J^0 =& \frac{1}{N} \vec{A}\cdot\vec{E}
\\
J^i =& \chi E^i - \frac{\vec{A} \cdot \vec{E}}{N} N^i 
-   \epsilon^{ijk} A_j B_k 
\label{J_idev}
\end{align}
so that the 3+1 form of  \eqref{NT_A_action} is 
\begin{widetext}
\begin{align}
S =&  \int d^4x \frac{N\sqrt{q}}{16\pi \Gt}\bigg\{ \Rcal + |K^2| - |K|^2  - 2 \Lambda
 + \KB \left( |\vec{E}|^2 -  |\vec{B}|^2 \right)
+ 2(2-\KB) \sigma \vec{A}\cdot\vec{E} 
+ 2  (2-\KB) \left(\chi \vec{E} -  \vec{A}\times \vec{B}  \right) \cdot \vec{D}\phi 
\nonumber
\\
&
+ \nu \left(\chi \sigma  + \vec{A}\cdot \vec{D}\phi \right)^{2} 
- \left(2-\KB +  \mu\right)\left[|\vec{A}|^2 \sigma^2 + 2\chi \sigma \vec{A} \cdot  \vec{D}\phi + |\vec{D}\phi|^2 + \left(\vec{A}  \cdot \vec{D}\phi\right)^2 \right]
 - \Ucal(\nu,\mu)
\bigg\}  + S_m[g]
\label{NT_A_action_3p1}
\end{align}
\end{widetext}
where $\Rcal$ is the Ricci scalar corresponding to the spatial metric $q_{ij}$,
and 
 $S$ is a functional of $(q_{ij},A_{i},\phi,\mu,\nu,N,N^{i})$, 
\section{Hamiltonian formulation}
\label{Hamiltonian_analysis}
Having cast the theory into a $3+1$ form, we now proceed to determine its Hamiltonian formulation, following the standard Dirac-Bergman
 procedure~\cite{Bergmann:1949zz,Anderson:1951ta,Dirac:1950pj,Dirac:1958sc,dirac2001lectures,Henneaux:1992ig} of constraint Hamiltonian systems.
 
The first step in passing to the Hamiltonian formulation is to determine the canonical momenta which are
\begin{align}
\Pi^{ij} \equiv& \frac{\delta S}{\delta \dot{q}_{ij}} = \frac{\sqrt{q}}{16\pi \Gt}\left( K^{ij}  - K q^{ij} \right)
\label{P_ij_def}
\\
\Pi^i \equiv& \frac{\delta S}{\delta \dot{A}_i}   =  \frac{\sqrt{q}}{8\pi \Gt}\bigg[ \KB  E^i + (2-\KB) \left(\sigma A^i
 +   \chi D^i\phi \right) \bigg] 
\label{P_i_def}
\\
\Pi \equiv& \frac{\delta S}{\delta \dot{\phi}}  =   \frac{\sqrt{q}}{8\pi \Gt}\bigg[
 (2-\KB)  \vec{A}\cdot\vec{E} 
+ \nu  \sigma  
\nonumber
\\
&
- \left( 2-\KB + \mu-\nu \right) \vec{A} \cdot \left( \sigma \vec{A} + \chi \vec{D}\phi \right)
\bigg]
\label{P_def}
\end{align}
Letting $\hat{\Pi} \equiv \Pi^{ij} q_{ij}$ the inverse relations are
\begin{align}
  K^{ij} =& \frac{16\pi \Gt}{\sqrt{q}} \left(\Pi^{ij}  - \frac{1}{2} \hat{\Pi}  q^{ij} \right)
\label{K_inv}
\\
\Xi \sigma
 =& \frac{8\pi \Gt}{\sqrt{q}}  \left[\Pi - \frac{2-\KB}{\KB}  \vec{A}\cdot \vec{\Pi}\right] 
\nonumber
\\
&
+   \left( 2 \frac{2-\KB}{\KB} + \mu-\nu \right)  \chi \vec{A} \cdot \vec{D}\phi
\label{sigma_inv}
\\
    \KB  E^i =& \frac{8\pi \Gt}{\sqrt{q}}  \Pi^i - (2-\KB) \left(\sigma  A^i +  \chi  D^i\phi  \right)
\label{E_inv}
\end{align}
where
\begin{equation}
\Xi = \chi^2\nu -\left(2 \frac{2-\KB}{\KB} + \mu\right)|\vec{A}|^2
\label{Xi_def}
\end{equation}
while the canonical momenta for $\mu$ and $\nu$ are identically zero:
\begin{align}
\Pmu \equiv& \frac{\delta S}{\delta \dot{\mu}} \approx 0
\label{Pmu}
\\
\Pnu \equiv& \frac{\delta S}{\delta \dot{\nu}} \approx 0
\label{Pnu}
\end{align}

Using \eqref{K_inv},   \eqref{sigma_inv} and \eqref{E_inv} to remove $K_{ij}$, $E_i$ and $\sigma$ from \eqref{NT_A_action_3p1} leads to the Hamiltonian form of the action,
\begin{align}
S &= \int d^4x \bigg\{ \Pi^{ij}\dot{q}_{ij} + \Pi^{i}\dot{A}_{i} + \Pi \dot{\phi} + \Pmu\dot{\mu} + \Pnu\dot{\nu}\nn\\
	& - N{\cal H} - N^{i}{\cal H}_{i} -  \lambdamu \Pmu - \lambdanu  \Pnu
\bigg\},
\label{HamiltonianAction}
\end{align}
where we have added Lagrange multipliers $\lambdamu$ and $\lambdanu$ imposing the constraints \eqref{Pmu} and \eqref{Pnu} and where
\begin{align}
\Hcal_i =& - 2D_j \Pi^j_{\;\;i} + \Pi  D_i\phi -  \vec{D}\cdot \vec{\Pi}   \; A_i - \epsilon_{ijk}  \Pi^j B^k 
\nn\\
&
+ \Pi^{(\mu)}D_{i}\mu + \Pi^{(\nu)}D_{i}\nu 
\label{Hcal_i}
\end{align}
is the  diffeomorphism constraint
and
\begin{widetext}
\begin{align}
\Hcal =&
 \frac{8\pi \Gt}{\sqrt{q}}\left[  2\Pi^{ij} \Pi_{ij}  -  \hat{\Pi}^2 + \frac{1}{2\KB}  |\vec{\Pi}|^2 + \frac{C_1^2}{2\Xi} \right]
+ \chi \left[
   \frac{C_1 C_2}{\Xi} \vec{A} \cdot \vec{D}\phi
+   \vec{D} \cdot \vec{\Pi}
-  \frac{2-\KB}{\KB}  \vec{\Pi} \cdot    \vec{D}\phi 
\right]
+ \frac{\sqrt{q}}{16\pi \Gt}\bigg\{ 
-  \Rcal  + 2 \Lambda 
+ \KB  |\vec{B}|^2 
\nonumber
\\
&
+ \left[ \frac{C_2^2\chi^2}{\Xi}  + 2-\KB + \mu-\nu \right] \left[\vec{A}\cdot \vec{D}\phi \right]^{2} 
+  2  (2-\KB) \vec{A}\times \vec{B}  \cdot \vec{D}\phi 
+ \left[ 2-\KB +  \mu
 + \frac{(2-\KB)^2}{\KB}  \chi^2 
 \right] |\vec{D}\phi|^2 
+ \Ucal
\bigg\},
\label{Hcal}
\end{align}
\end{widetext}
the Hamiltonian constraint. We have defined
\begin{align}
C_1 \equiv \Pi - \frac{2-\KB}{\KB}  \vec{A}\cdot \vec{\Pi},
\label{C_1_def}
\\
C_2 \equiv  2 \frac{2-\KB}{\KB} +\mu-\nu.
\label{C_2_def}
\end{align}
Setting the additional fields and their canonical momenta in \eqref{Hcal_i} and \eqref{Hcal} to zero, one recovers the equivalent constraints in GR~\footnote{
That is,
\begin{equation}
\Hcal_i^{(GR)} = - 2D_j \Pi^j_{\;\;i},
\end{equation}
 and 
\begin{equation}
\Hcal^{(GR)} = \frac{8\pi \Gt}{\sqrt{q}}\left[  2\Pi^{ij} \Pi_{ij}  -  \hat{\Pi}^2 \right]
+ \frac{\sqrt{q}}{16\pi \Gt}\left( -  \Rcal  + 2 \Lambda \right)
.
\end{equation}
}.
Note that to arrive at the action (\ref{HamiltonianAction}) we have defined the coefficients multiplying $(\Pi^{(\mu)},\Pi^{(\nu)})$ to be $(\lambda^{(\mu)}+N^{i}D_{i}\mu,\lambda^{(\nu)}+N^{i}D_{i}\nu)$ 
which we are free to do at the beginning of the constraint analysis. This leads to the 2nd line in \eqref{Hcal_i}.

The combination 
\begin{equation}
\HcalP =  N\Hcal  + N^{i}\Hcal_{i} + \lambdamu\Pmu + \lambdanu \Pnu 
\label{Hcal_T}
\end{equation}
 is the \emph{primary} Hamiltonian density and it is a sum of constraints on the phase space which is coordinatized 
by $\{q_{ij},\Pi^{ij},A_{i},\Pi^{i},\varphi,\Pi,\mu,\Pmu,\nu,\Pnu\}$. These constraints are obtained by varying \eqref{HamiltonianAction} 
with $\{N,N^{i},\lambdamu,\lambdanu\}$ and are given respectively by:
\begin{subequations}
\begin{align}
\Hcal &\approx 0 \\
\Hcal_{i} &\approx 0 \\
\Pmu &\approx 0 \\
\Pnu &\approx 0
\end{align}
\label{Primary_Constraints}
\end{subequations}
where $\approx$ denotes that an equation `weakly vanishes' which means that the equation holds on the submanifold of phase space defined by the constraints but need not hold in regions of phase space not on the constraint submanifold \cite{dirac2001lectures}. This can occur, for example, if a phase space function $f$ is equal to a combination of the constraints themselves.

The next step is to check whether these constraints are preserved by the time evolution generated by $\HcalP$.

\section{The propagation of constraints}
\label{Constraint_analysis}

\subsection{Poisson brackets}
For a phase space coordinatized by fields $Q_{I}(t,\coordT{x})$ and $P_{I}(t,\coordT{x})$ (where here indices $I,J,\dots$ label the different fields), 
it is useful to introduce the Poisson bracket defined for quantities $A(Q_{I},P_{J})$, $B(Q_{I},P_{J})$.
If $\tau_{i\ldots j}^{k\ldots l}(x)$ is a general tensor field and $\Fcal[\tau_{i\ldots j}^{k\ldots l}(x)]$  
 a functional of $\tau_{i\ldots j}^{k\ldots l}(x)$, then the functional derivative of  $\Fcal[\tau_{i\ldots j}^{k\ldots l}(x)]$   wrt $\tau_{i\ldots j}^{k\ldots l}(x)$,
in three dimensions   is defined as
\begin{align}
        \frac{\delta \Fcal[\tau_{i\ldots j}^{k\ldots l}(\coordT{x})]}{\delta \tau_{a\ldots b}^{c\ldots d}(\coordT{y})} &=
 \lim_{\epsilon \rightarrow 0} \frac{1}{\eps}\bigg\{
\Fcal\big[ \tau_{i\ldots j}^{k\ldots l}(\coordT{x})  \nn\\
&+ \eps \delta^{(3)}(\coordT{y}-\coordT{x})\delta^{k}_{S\{c}\dots \delta^{l}_{d\}}\delta^{S\{a}_{i}\dots \delta^{b\}}_{j} \big]
\nonumber
\\
&       
-\Fcal\left[\tau_{i\ldots j}^{k\ldots l}(\coordT{x})\right]
\bigg\}
\end{align}
with $\delta^{(3)}(\coordT{y}-\coordT{x})$ being the three-dimensional Dirac delta-function and $S\{ab..cd\}$
applies the symmetries of the tensor field $\tau$ e.g. if $\tau^{ijkl}=\tau^{jikl}$, $\tau^{ijkl}=-\tau^{ijlk}$ then $S\{ijkl\}= (ij)[kl]$. 
With this definition, the Poisson bracket of  $A(Q_{I},P_{J})$ and $B(Q_{I},P_{J})$ is defined as
\begin{align}
\Poisson{A}{B} \equiv \sum_{I}\int d^3x \left(\frac{\delta A}{\delta Q_{I}}\frac{\delta B}{\delta P_{I}} - \frac{\delta A}{\delta P_{I}}\frac{\delta B}{\delta Q_{I}} \right)
\label{def_Poisson}
\end{align}
For time evolution according to any general Hamiltonian  $H$ (corresponding to general Hamiltonian density $\mathscr{H}$)
\begin{equation}
H = \int d^3x \mathscr{H}
\label{Hany}
\end{equation} 
 we have Hamilton's equations for quantities $f(Q_{I},P^{I})$ on phase space
\begin{align}
	\dot{f} &=   \Poisson{f}{H}.
\label{eq_f_dot}
\end{align}
In our case, we test the time evolution of the constraints according to $\Hpri$, that is,
letting $\Constr_I$ being any of the constraints in the set $\{\Hcal,\Hcal_i,\Pmu,\Pnu\}$, we require that $\dot{\Constr}_I \approx 0$ with $H \rightarrow \Hpri$ in \eqref{Hany}
and \eqref{eq_f_dot}.

It is generally more straightforward to evaluate the Poisson bracket of \emph{smeared} constraints i.e. we define, for some `test' function $N(\coordT{x})$:
\begin{align}
	\smear{\Hcal}{N}  \equiv \int d^{3}y N(\coordT{y}) \; \Hcal[Q_{I}(\coordT{y}),P^{I}(\coordT{y})].
\end{align}
With the above definition, \eqref{Hcal_T} is rewritten as
\begin{equation}
\Hpri =  \smear{\Hcal}{N} + \smear{\Hcal_i}{N^i} +  \smear{\Pmu}{\lambdamu} +  \smear{\Pnu}{\lambdanu} .
\label{Hpri_smeared}
\end{equation}

To proceed with the evaluation of Poisson brackets, it is useful to have at hand the following two results.
If $\Hcal$ depends algebraically on $Q_{I}$ then $ \frac{\delta \Hcal[N]}{\delta Q_{I}(\coordT{x})} =  N(\coordT{x}) \frac{\partial \Hcal}{\partial Q_{I}}(\coordT{x})$. Furthermore,
given arbitrary tensor fields $\sigma^{i\ldots j}_{k\ldots l}$ and $\tau_{i\ldots j}^{k\ldots l}$ then
\begin{align}
 &\frac{\delta}{\delta \tau_{b_1\ldots b_s}^{a_1\ldots a_r}} \int d^3y \, \sigma^{d_1\ldots d_s}_{c_1\ldots c_r} \, \Lie{\vec{N}}\tau_{d_1\ldots d_s}^{c_1\ldots c_r}
\nn\\
&= - \Lie{\vec{N}}  \sigma^{d_1\ldots d_s}_{c_1\ldots c_r}\delta_{S\{a_1}^{c_1}\ldots\delta_{a_r\}}^{c_r}\delta^{S\{b_1}_{d_1}\ldots\delta^{b_s\}}_{d_s}
\end{align}
where $\Lie{\vec{N}}$ is the Lie derivative according to the vector field $\vec{N}$.

To satisfy \eqref{eq_f_dot} there are ten Poisson brackets to be evaluated among the smeared constraints $\Constr_I$. Out of these, three are trivial, namely,
$\Poisson{\Pmu}{\Pmu} = \Poisson{\Pnu}{\Pnu}=\Poisson{\Pmu}{\Pnu}=0$ vanish strongly.  Let us now evaluate the remaining seven.

\subsection{Poisson brackets involving a diffeomorphism constraint}
Starting from \eqref{Hcal_i} it can be shown that
\begin{align}
	\smear{\vec{\Hcal}}{\vec{N}} 	& \overset{b}{=} \int d^3x\big(
  \Pi^{ij} \Lie{\vec{N}}q_{ij}
+ \Pi \Lie{\vec{N}}\phi
+  \Pi^i \Lie{\vec{N}}A_i\nn\\
&+\Pi^{(\mu)}\Lie{\vec{N}}\mu+\Pi^{(\nu)}\Lie{\vec{N}}\nu\big) 
\end{align}
where $\overset{b}{=}$ means equal to up to a boundary term. 
As in \cite{Romano1991}, given a general \emph{smeared tensor density}  $ \smear{\tau_{i\ldots j}^{k\ldots l}}{\sigma^{i\ldots j}_{k\ldots l}}$  one finds
\begin{align}
\Poisson{ \smear{\vec{\Hcal}}{\vec{N}}  }{ \smear{\tau_{i\ldots j}^{k\ldots l}}{\sigma^{i\ldots j}_{k\ldots l}}}
 =& 
\smear{\tau_{i\ldots j}^{k\ldots l}}{\,\Lie{\vec{N}} \,\sigma^{i\ldots j}_{k\ldots l}\,} \label{diff_tau}
\end{align}
and choosing $\sigma \rightarrow N^i$ and  $\tau \rightarrow \Hcal_i$ we find the Poisson bracket between two diffeomorphism constraints as
\begin{align}
\Poisson{ \smear{\vec{\Hcal}}{\vec{N}}  }{ \smear{\vec{\Hcal}}{\vec{N}'}}  
 =&
\smear{\vec{\Hcal}}{\Lie{\vec{N}}\vec{N}'}  
\end{align}
Similarly choosing $\sigma\rightarrow N$ and $\tau \rightarrow \Hcal$ we find the  Poisson bracket between the diffeomorphism and the Hamiltonian 
constraint as
\begin{align}
\Poisson{ \smear{\vec{\Hcal}}{\vec{N}}  }{ \smear{\Hcal}{N}}  =& \smear{\Hcal}{\Lie{\vec{N}} N} 
\end{align}
These two Poisson brackets are identical to the ones for GR, however, the phase-space is now enlarged due to the additional fields $A_i$, $\phi$, $\mu$, and $\nu$.

Lastly, setting $\sigma\rightarrow \lambdaA$ and $\tau \rightarrow \Pi^{(A)}$ (with $A$ denoting either $\mu$ or $\nu$) trivially gives
 $\Poisson{ \smear{\vec{\Hcal}}{\vec{N}}  }{ \smear{\Pmu}{\lambdamu}} = \smear{\Pmu}{\Lie{\vec{N}}\lambdamu }$  and 
$\Poisson{ \smear{\vec{\Hcal}}{\vec{N}}  }{ \smear{\Pnu}{\lambdanu}} = \smear{\Pnu}{\Lie{\vec{N}}\lambdanu }$  which vanish weakly. 

\subsection{Poisson brackets of two Hamiltonian constraints}
Now consider the Poisson bracket of two Hamiltonian constraints, that is $\Poisson{\smear{\Hcal}{N}}{\smear{\Hcal}{N'}}$.  We let
\begin{align}
\frac{\delta \smear{\Hcal}{N} }{ \delta Q_I} =&  N \Acal_{(Q_I)} +  \Acal_{(Q_I)}^i D_i N
\nonumber
\\
&
+ \frac{\sqrt{q}}{16\pi \Gt} \delta_{(Q_I, q_{ij})}\left[ q^{ij}  \vec{D}^2 N - D^i D^j N  \right]
\label{Ham_Q_I}
\\
\frac{\delta \smear{\Hcal}{N} }{ \delta P_I} =& N \Bcal_{(P_I)} +  \Bcal_{(P_I)}^i D_i N
\label{Ham_P_I}
\end{align}
where $Q_I = \{ q_{ij}, A_i, \phi\}$, $P_I = \{ \Pi^{ij}, \Pi^i, \Pi\}$,
 $\delta_{(Q_I, q_{ij})} = 1$ if $Q_I \rightarrow q_{ij}$ and zero otherwise, and the exact form of $\Acal_{(Q_I)}$, $\Acal_{(Q_I)}^i$, $\Bcal_{(P_I)} $ and $\Bcal_{(P_I)}^i$ is 
displayed in appendix \ref{AB_coeff}.

Using equation \eqref{def_Poisson} directly and making use of \eqref{Ham_Q_I} and \eqref{Ham_P_I}, the cross terms proportional to $NN'$ vanish due to the antisymmetry of the Poisson bracket.
In addition, from  appendix \ref{AB_coeff} we have that
 $ \Acal_{(q_{ij})}^k  = \Bcal^i_{(\Pi^{jk})} = \Bcal^i_{(\Pi)} = 0$ while $\Acal_{(A_j)}^i = -\Acal_{(A_i)}^j$ and  $\Bcal^i_{(\Pi^{j})} =  
 -\chi  \delta^i_{\;\;j}$, 
so that after some integrations by parts we find
\begin{align}
 \Poisson{\smear{\Hcal}{N}}{\smear{\Hcal}{N'}} &= \int d^3x 
\bigg\{
 \chi  \left[ D_j \Acal_{(A_k)}^j -\Acal_{(A_k)}  \right]
\nonumber
\\
&
 \ +\frac{\sqrt{q}}{16\pi \Gt} \left[ q^{ij} D^k  \Bcal_{(\Pi^{ij})} - q^{jk} D^i \Bcal_{(\Pi^{ij})}  \right]
\nonumber
\\
&
\ -   \Acal_{(A_i)}^k\left[ D_i  \chi  +  \Bcal_{(\Pi^i)} \right]
\nonumber
\\
&
\  -   \Bcal_{(\Pi)}  \Acal_{(\phi)}^k 
\bigg\} \left( N D_k N'  - N' D_k N    \right)
\end{align}
Finally, plugging in all the expressions from appendix \ref{AB_coeff}  leads to
\begin{align}
 \Poisson{\smear{\Hcal}{N}}{\smear{\Hcal}{N'}} &= \int d^3x 
\bigg[ 
 -2 D_i  \Pi^{ik}
+ \Pi \vec{D}^k\phi
-   A^k \vec{D} \cdot \vec{\Pi} 
\nonumber
\\
&
+  \Pi^i D^k A_i
-  \Pi^i D_i A^k
\bigg] \left( N D_k N'  - N' D_k N    \right)
\end{align}
so that after integrations by parts we get
\begin{align}
\Poisson{\smear{\Hcal}{N}}{\smear{\Hcal}{N'}} \approx \smear{\Hcal_i}{N D^iN' -N'D^i N} \label{H_H_Hi}
\end{align}
Once again, this Poisson bracket is identical to the ones for GR albeit for an enlarged phase-space. In this regard, a comment is in order. When a theory with spacetime diffeomorphism symmetry is cast 
into Hamiltonian form, there generally appear constraints corresponding respectively to the generators of spatial diffeomorphisms and time reparameterizations. 
These constraints will generally \emph{weakly} obey the algebra found above, known as the Dirac hypersurface
 deformation algebra (or appropriate sub-algebras following partial spacetime gauge fixing) and it can be shown to hold in other extensions of GR, such as, examples of Horndeski 
and beyond Horndeski theories \cite{Lin:2014jga,Langlois:2015skt}.

\subsection{Secondary constraints}
Now consider the Poisson bracket of $\smear{\Hcal}{N}$ with $\smear{\Pmu}{\lambdamu}$ and $\smear{\Pnu}{\lambdanu}$ leading to
\begin{align}
 \Poisson{\smear{\Hcal}{N}}{\smear{\Pmu}{\lambdamu}}  &=  \int d^{3}x N \lambdamu  \frac{\partial {\Hcal}}{\partial \mu}
\end{align}
and 
\begin{align}
 \Poisson{\smear{\Hcal}{N}}{\smear{\Pnu}{\lambdanu}}  &=  \int d^{3}x N \lambdanu  \frac{\partial {\Hcal}}{\partial \nu}
\end{align}
respectively. Hence, for general $N$, $\lambdamu$ and $\lambdanu$, 
the requirement these Poisson brackets vanish weakly implies the additional secondary constraints on phase space,
\begin{align}
\Smu  \equiv   \frac{\partial {\Hcal}}{\partial \mu} =  \frac{\sqrt{q}}{16\pi\Gt} \left( \Ycal +  \frac{\partial \Ucal}{\partial \mu} \right) &\approx 0
\label{sec1}
\end{align}
and
\begin{align}
 \Snu  \equiv \frac{\partial {\Hcal}}{\partial \nu} =  \frac{\sqrt{q}}{16\pi \Gt}\left( -  \Qcal^2 + \frac{\partial\Ucal}{\partial\nu} \right)  &\approx 0 
\label{sec2}
\end{align}
where remember that $\Xi=\Xi(\mu,\nu)$ from \eqref{Xi_def}. 
Equations \eqref{sec1} and \eqref{sec2} correspond to the Euler-Lagrange equations for the auxiliary fields $\mu$ and $\nu$.
In other words, given a prescribed $\Ucal(\mu,\nu)$ one can use  \eqref{sec1} and \eqref{sec2}  to determine $\mu(\Qcal,\Ycal)$ and $\nu(\Qcal,\Ycal)$, 
where $\Qcal$ and $\Ycal$ are to be evaluated in phase space.
Using \eqref{sigma_inv} followed by \eqref{eq_Ycal} and \eqref{eq_Qcal} to collect terms together, 
one finds that 
\begin{align}
\Qcal &= 
 \frac{1}{\Xi} \left[
   \left(2 \frac{2-\KB}{\KB} + \mu  \right)   \vec{A} \cdot \vec{D}\phi
  + \frac{8\pi \Gt}{\sqrt{q}} \chi C_1
\right]
 \label{qhamcon}
\end{align}
and
\begin{align}
\Ycal &= 
 |\vec{D}\phi|^2 
+  \left(\vec{A} \cdot \vec{D}\phi\right)^2
+ \frac{|\vec{A}|^2}{\Xi^2} \left[  \frac{8\pi \Gt}{\sqrt{q}} C_1 + C_2 \chi \vec{A} \cdot \vec{D}\phi \right]^2
\nonumber
\\
&
+ \frac{2\chi}{\Xi} \left(  \frac{8\pi \Gt}{\sqrt{q}} C_1 + C_2 \chi \vec{A} \cdot \vec{D}\phi \right)  \vec{A} \cdot \vec{D}\phi,
\label{yhamcon}
\end{align}
respectively. This procedure then reconstructs $\Fcal(\Qcal,\Ycal)$ through \eqref{Fcal_to_Ucal} in terms of phase space variables with the help of \eqref{qhamcon} and \eqref{yhamcon}.

\begin{table}
 \begin{tabular}{|c|c|c|c|c|c|c|}
\hline
 & \multicolumn{4}{c|}{Primary}  
 & \multicolumn{2}{c|}{Secondary} 
\\
\hline
             &    $\Hcal_i$    &  $\Hcal$   & $\Pmu$ & $\Pnu$ & $\Smu$ & $\Snu$  \\
\hline
   $\Hcal_i$ &     $\Hcal_i$  &  $\Hcal$     &   $\Pmu$  &  $\Pnu$   &    $\Smu$     &   $\Snu$    \\
   $\Hcal$   &               &   $\Hcal_i$     & $\Smu$ & $\Snu$ &  \lagrange{U^{(\mu)}} &   \lagrange{U^{(\nu)}}      \\ 
   $\Pmu$    &               &              &   $0$  &  $0$   &  \lagrange{C^{(\mu)(\mu)}}  & \lagrange{C^{(\nu)(\mu)}}   \\
   $\Pnu$    &               &              &        &  $0$   &  \lagrange{C^{(\mu)(\nu)}}     &  \lagrange{C^{(\nu)(\nu)} }         \\
\hline
   $\Snu$    &               &              &        &        &     \lagrange{E^{(\mu)(\mu)}}    &      \lagrange{E^{(\mu)(\nu)}}    \\
   $\Smu$    &               &              &        &        &        &     \lagrange{E^{(\nu)(\nu)}}
\\
\hline
 & \multicolumn{1}{c|}{First Class}
 & \multicolumn{5}{c|}{Second Class}
\\
\hline
\end{tabular}
\caption{Table of constraints and their Poisson Brackets, showing whether they vanish strongly, weakly, or not at all. The classification into primary/secondary and first/second class is marked.
Note also that the combination $\Htot$ defined through \eqref{Htot} is first class, with $\lambdaA$ being functions of all the phase space variables as determined through \eqref{find_lambdaB}, 
 even though some individual parts  of $\Htot$ are second class.
 }
\label{tab_constraints}
\end{table}

Having found the secondary constraints \eqref{sec1} and \eqref{sec2}, the analysis is not necessarily finished. 
We now define the secondary Hamiltonian through
\begin{equation}
\Hsec = \Hpri + \int d^3x \left[ \umu \Smu + \unu \Snu \right]
\end{equation}
where $\umu$ and $\unu$ are Lagrange multipliers enforcing the secondary constraints $\Smu$ and $\Snu$ respectively. We then check that 
all constraints (primary and secondary) are preserved in time by taking their Poisson bracket with $\Hsec$. 
Since the Poisson bracket of all constraints with $\Hpri$ is weakly vanishing by default, it is sufficient to consider the Poisson brackets of $\Smu$ and $\Snu$ with
any constraint in the set $\{\Hcal_i, \Hcal, \Pmu, \Pnu, \Smu,\Snu\}$, implying eleven brackets in total.

We set the index $A =  \{ \mu, \nu\}$ and consider collectively the vector $\smear{\SA}{\uA}$. From equation (\ref{diff_tau}) we have:
\begin{align}
 \Poisson{\smear{\SA}{\uA}}{\smear{\vec{\Hcal}}{\vec{N}}} &=  - \smear{\SA}{ \Lie{\vec{N}} \uA }
\end{align}
which therefore vanish weakly. Hence, $\vec{\Hcal}$ remains a first class constraint.
The remaining nine brackets do not vanish, meaning that all other constraints are second class.  Consider first 
\begin{align} 
\partial_t \smear{\PA}{\lambdaA}  
 \approx \sum_B \Poisson{\smear{\PA}{\lambdaA}}{ \smear{ \SB }{ \uB } } \approx 0.
\end{align}
The involved Poisson brackets are evaluated as
\begin{align} 
 \Poisson{\smear{\PA}{\lambdaA}}{ \smear{ \SB }{ \uB } }  = - \smear{C^{AB} }{\lambdaA \uB}
\label{eq_PA_SB}
\end{align}
where 
\begin{align}
C^{AB} \equiv \frac{\partial \SB }{\partial A} = 
\begin{pmatrix}                                                                                                                                                                            
\frac{\partial \Smu}{\partial \mu}  &  \frac{\partial \Smu}{\partial \nu}
 \\
  \frac{\partial \Snu}{\partial \mu}   &  \frac{\partial \Snu}{\partial \nu}
        \end{pmatrix}
\end{align}
Then \eqref{eq_PA_SB} gives two homogeneous equations for two unknowns, the Lagrange multipliers $\umu$ and $\unu$, and therefore implies that they must both vanish. 

Consider now \begin{align}
\partial_t \smear{\Hcal}{N}
  \approx \sum_B \Poisson{\smear{\Hcal}{N}}{ \smear{ \SB }{ \uB } } \approx 0.
\end{align}
The involved Poisson brackets are evaluated as
\begin{equation}
 \smear{U^B}{\xiB} \equiv \Poisson{\smear{\Hcal}{N}}{ \smear{ \SB }{ \xiB } }  =   \Poisson{\smear{\Hcal}{N}}{ \frac{\partial  \smear{ \Hcal }{ \xiB }  }{\partial B} } 
\label{eq_Hcal_SB}
\end{equation}
for arbitrary functions $\xiA$ (not necessarily the Lagrange multipliers $\uA$),
 which then allows us to use the $\Acal$ and $\Bcal$ coefficients defined in \eqref{Ham_Q_I} and \eqref{Ham_P_I} and displayed in appendix \ref{AB_coeff}.
Explicitly, after some integrations by parts and further computations we find
\begin{align}
 U^A &=  
N \bigg(
 -\Bcal_{(\Pi^{ij})} \frac{\partial  \Acal_{(q_{ij})}  }{\partial A}
 + \Acal_{(A_i)}  \frac{\partial \Bcal_{(\Pi^i)}}{\partial A}
-  \Bcal_{(\Pi^i)} \frac{\partial \Acal_{(A_i)} }{\partial A}
\nonumber
\\
&
+  \Acal_{(\phi)} \frac{\partial \Bcal_{(\Pi)}}{\partial A} 
+  D_i \Bcal_{(\Pi)} \frac{\partial  \Acal^i_{(\phi)}  }{\partial A}
\bigg).
\label{eq_U_A}
\end{align} 
It is crucial to note that $U^{(A)}$ can vanish neither strongly nor weakly. This can be easily seen by focussing on the
$\Bcal_{(\Pi^{ij})} \frac{\partial  \Acal_{(q_{ij})}  }{\partial A}$ term, which is the only term containing $\Pi^{ij}$ and thus cannot be cancelled by other terms.
Back to \eqref{eq_Hcal_SB}, we set $\xiB \rightarrow \uB \approx 0$ which results in $ \partial_t \smear{\Hcal}{N} =  \smear{U^{(\mu)}}{\umu}  +  \smear{U^{(\nu)}}{\unu} \approx 0$ without requiring any additional constraints
or equations involving the Lagrange multipliers.

Lastly, consider  $\partial_t \smear{\SA}{\xiA}$. Using our definitions from \eqref{eq_PA_SB} and \eqref{eq_Hcal_SB}, we have
\begin{align}
\partial_t \smear{\SA}{\xiA} =&
 - \smear{U^{A}}{\xiA}
 - \sum_B \smear{C^{AB}}{\xiA \lambdaB}
\nonumber
\\
&
+ \sum_B \Poisson{\smear{\SA}{\xiA}}{\smear{\SB}{\uB}}
\end{align}
Now, the last Poisson bracket depends linearly on $\uB$ (and its derivative) and the same for $\xiA$, hence, it vanishes when $\uB \approx 0$. Require then that
$\partial_t \smear{\SA}{\xiA} \approx 0$ leads to the linear system of equations
\begin{align}
  U^{A} + \sum_B C^{AB} \lambdaB \approx 0
\label{find_lambdaB}
\end{align}
%
If the matrix $C^{AB}$ is invertible then the two equations above \emph{determine} the Lagrange multipliers $\lambda^{A}$ which then become functions of all the phase-space variables.
 Then the Hamiltonian analysis is complete and no further constraints in phase-space are required, with the conclusion that the theory possesses 
three first class primary constraints $\Hcal_i$, 
 three second class primary constraints  $\Hcal$ and $\PA$, and two second class secondary constraints  $\SA$. We list all the constraints and their Poisson brackets 
in table \ref{tab_constraints}.
On the other hand, if the matrix has a vanishing determinant, there exists a left null eigenvector which when applied 
to the consistency relation may produce further constraints, called \emph{tertiary} constraints. Given the forms
of (\ref{qhamcon}),(\ref{yhamcon}) and equations of motion (\ref{sec1}),(\ref{sec2}) it is to be expected that $C^{AB}$ is indeed invertible in general situations.

%

\subsection{Hamiltonian evolution}
Having found all the constraints and solved for the Lagrange multipliers, we may now form the \emph{first class} Hamiltonian $\Htot$. We find this as the secondary Hamiltonian with the 
Lagrange multiplies subbed-in. Given that $\uA\approx 0$, we have that $\Hsec \approx \Hpri$, hence,
\begin{equation}
\Htot = \int d^3x \left[ N \Hcal + N^i \Hcal_i + \lambdamu \Pmu + \lambdanu \Pnu \right]
\label{Htot}
\end{equation} 
with $\lambdamu$ and $\lambdanu$ being functions of the phase-space variables $\{q_{ij},A_i,\phi,\Pi^{ij},\Pi^i,\Pi,\mu,\nu\}$, that is, $\Pmu$ and $\Pnu$ are absent from
$\lambdaA$. Hamilton's equations are then found using \eqref{eq_f_dot} with $H\rightarrow\Htot$. We find
\begin{subequations}
\begin{align}
\dot{q}_{ij} \approx& N  \Bcal_{(\Pi^{ij})} + 2 D_{(i} N_{j)}
\label{q_ij_dot}
\\
\dot{\Pi}^{ij} \approx& -N  \Acal_{(q_{ij})} - 2\Pi^{k(j} D_k N^{i)} + D_k\left( N^k\Pi^{ij} \right) 
\nonumber
\\
&
+ \frac{\sqrt{q}}{16\pi \Gt}\left(  D^i D^j N    -q^{ij}  \vec{D}^2 N \right)
\label{Pi_ij_dot}
\\
\dot{A}_{i}\approx& N  \Bcal_{(\Pi^{i})} - \chi D_i N + N^j D_j A_i + A_j D_i N^j
\label{A_i_dot}
\\
\dot{\Pi}^i \approx & - N  \Acal_{(A_i)} - \Acal^j_{(A_i)}  D_j N  + D_k\left( N^k  \Pi^i \right) - \Pi^j D_j N^i
\label{Pi_i_dot}
\\
\dot{\phi} \approx & N \Bcal_{(\Pi)}  +  N^i D_i \phi
\label{phi_dot}
\\
\dot{\Pi} \approx &  D_i \left( \Pi N^i    - N \Acal^i_{(\phi)}  \right)
\label{Pi_dot}
\\
 \dot{\mu} \approx& \lambdamu + N^{i}D_{i}\mu
\label{mu_dot}
\\
 \dot{\nu} \approx& \lambdanu + N^{i}D_{i}\nu
\label{nu_dot}
\end{align}
\label{Hamilton_eq}
\end{subequations}
along with the constraints \eqref{Primary_Constraints}, \eqref{sec1} and \eqref{sec2}, and where to reiterate, $\lambdaA$ are functions of all the phase-space variables 
obtained after solving \eqref{find_lambdaB}.

\section{An example: Small perturbations around Minkowski spacetime}
\label{Minkowski_example}
The  Hamiltonian density is rather complicated in general so for illustration it is useful to have a look at a simple example. 
We consider a background Minkowski spacetime and the linear evolution of small perturbations to this background.

\subsection{Background Minkowski solution}
Before discussing the choice of function $\Ucal(\nu,\mu)$, let us first determine what conditions a Minkowski background imposes on the fields, in the general sense.
Note that a cosmological constant $\Lambda$ can always be absorbed into the definition of $\Ucal(\nu,\mu)$, hence, we set $\Lambda=0$ without any loss of generality.
We use an overbar to denote the values that fields take in the background and we fix  the background gauge  to
\begin{align}
\bar{N} = 1 \qquad \bar{N}^{i} = 0. 
\end{align}
Moreover, for Minkowski spacetime we have $\bar{q}_{ij} = \delta_{ij}$ and then \eqref{q_ij_dot} trivially gives $\Pib^{ij} = 0$ (and so $\bar{\Bcal}_{(\Pi^{ij})}=0$). Now, the background solution should not
violate the isometries of Minkowski spacetime, including rotational invariance, hence, all three dimensional vector fields should vanish: 
\begin{align}
\bar{A}_i = \Pib^i = D_i \phib = 0
\label{no_vectors_on_Minkowski}
\end{align}
so that  \eqref{A_i_dot} and \eqref{Pi_i_dot} are trivially satisfied
(and   $\bar{\Bcal}_{(\Pi^{i})}=  \bar{\Acal}_{(A_i)}  =  \bar{\Acal}^i_{(A_j)}  =0$,
 while $\bar{\Bcal}^i_{(\Pi^{j})} =  -  \delta^i_{\;\;j}$). The conditions \eqref{no_vectors_on_Minkowski}
 in turn imply that $\bar{\chi} =1$  and $\phib=\phib(t)$ leading in addition to $\bar{\Ycal}=0$. Hence, from \eqref{Xi_def} we find 
 $\bar{\Xi} = \nub$ and from \eqref{C_1_def} we get $\bar{C}_1 = \Pib$. 

Meanwhile, \eqref{no_vectors_on_Minkowski} leads to $\bar{\Acal}^i_{(\phi)} = 0$ (and hence $\bar{\Acal}_{(\phi)} = 0$) so that 
from \eqref{eq_U_A} we find $\bar{U}^A = 0$ and hence, \eqref{find_lambdaB} leads to $\lambdaA = 0$. This last condition then implies
through \eqref{mu_dot} and \eqref{nu_dot}  that $\mub$ and $\nub$ are functions of $\vec{x}$ only.
Moreover, \eqref{Pi_dot} implies that  $\Pib$ is also a function of $\vec{x}$ only.

Now from \eqref{Pi_ij_dot} we have that $\bar{\Acal}_{(q_{ij})}  = 0$ and this imposes that
 $\bar{\Ucal} = (8\pi \Gt)^2 \Pib^2/\nub $. Finally, we have that $\bar{\Bcal}_{(\Pi)} = 8\pi \Gt \Pib / \nub$ is time-independent, so that
\eqref{phi_dot} may be integrated to get $\phib = \Qcal_0 t$, where $\Qcal_0 = 8\pi \Gt  \Pib / \nub$ is a constant.  We thus set
\begin{align}
  8\pi \Gt  \Pib =  \Qcal_0 \nub
\end{align}
without loss of generality, implying that $\Pib$ and $\nub$ have the same $\vec{x}$ dependence, and that $\Qcal_0$, being a constant,  is independent of $\nub$.

 Now \eqref{qhamcon} gives $\bar{\Qcal} = \Qcal_0$
so that the constraint \eqref{sec2} gives $\partial\Ucal/\partial\nub =  \Qcal_0^2$. Since we also have that 
 that $\partial\Ucal/\partial\mub =  0$ from \eqref{sec2}, we find that
\begin{align}
 \bar{\Ucal} = \Qcal_0^2 \nub
\end{align}
which is consistent with our discussion in the previous paragraph. Note that the above condition, is a condition that $\Ucal$ must satisfy in order to have Minkowski solutions
but it does not completely determine the general form of the function $\Ucal$.

Finally, before discussing function choices, let us connect with the original function $\Fcal$. From \eqref{Fcal_to_Ucal} we find the relations
\begin{align}
\mu = \frac{\partial \Fcal}{\partial \Ycal}
\end{align}
\begin{align}
\nu = -\frac{\partial \Fcal}{\partial \Qcal^2}
\end{align}
Hence, since $\bar{\Ycal} =0$ and $\bar{\Qcal} = \Qcal_0$ is a constant, this implies that $\Fcal$ and its derivatives will also be, at best, constants. We immediately get that
both $\mub$ and $\nub$ are constants and hence, so is $\Pib$. Incidentally, \eqref{Fcal_to_Ucal} returns $\bar{\Fcal} = 0$.

\subsection{Choice of function}
We now turn to the choice of function $\Ucal(\nu,\mu)$ in order to pave the way for departures from Minkowski.
Our goal here is to compare to \cite{SkordisZlosnik2021}, hence, we restrict ourselves to cases where the function $\Fcal(\Ycal,\Qcal)$ of (\ref{NT_A_action}) takes the form
\begin{align}
\Fcal(\Ycal,\Qcal) &=  (2-\KB) \lambdas \Ycal - 2\Kcal_{2}\left(\Qcal-\Qcal_0\right)^2   
\end{align}
where $\lambdas$ and $\Kcal_2$ are constants.
The above functional form is motivated  by making sure that at the weak field quasistatic limit Newtonian gravity is recovered and that 
the large-scale FLRW cosmology admits dust solutions. Exact Minkowski is recovered when $\Ycal=0$ and $\Qcal = \Qcal_0$ in accordance with the previous subsection.

We then find that
\begin{align}
\mu = (2-\KB) \lambdas
\label{eq_mu_lambdas}
\end{align}
and
\begin{align}
\nu =  2\Kcal_{2}\left(1 - \frac{\Qcal_0}{\Qcal}\right)
\end{align}
which inverts to
\begin{align}
\Qcal  =  \frac{\Qcal_0}{ 1- \frac{\nu}{2\Kcal_{2} }  }
\label{Qcal_nu}
\end{align}
so that 
\begin{align}
  \Ucal =&   \frac{\Qcal_0^2 \nu}{ 1- \frac{\nu}{2\Kcal_{2} }  } 
\end{align}
Using the above functional form into the constraint \eqref{sec1} leads to $\Ycal = 0$, while constraint \eqref{sec2} returns back
\eqref{Qcal_nu}. The last relation also leads to 
\begin{align}
\nub = 0,
\end{align}
and hence,
\begin{align}
\Pib = 0.
\end{align}

\subsection{Hamiltonian density to 2nd order}
We perturb  the lapse function as 
\begin{align}
        N =& 1 +  \Psi
\end{align}
where $\Psi$ is a small perturbation and let the shift $N_i = h_i$  be  a pure perturbation (zero background).
Additionally, we perturb our phase-space variables as
\begin{align}
	q_{ij} =& \delta_{ij} +  h_{ij} 
\\
\Pi^{ij} =& \varpi^{ij} 
\\
A_i =& \alpha_i
\\
\Pi^i =& \varpi^i 
\\
\phi =&  \Qcal_0 t + \varphi 
\\
\Pi =&  \frac{\Qcal_0}{8\pi \Gt} \nub + \varpi
\\
\nu =& \nub + \neta
\end{align}
where $h_{ij}$, $\alpha_i$,  $\varphi$, $\varpi^{ij}$, $\varpi^{i}$, $\varpi$  and $\neta$ are small perturbations,
and we further define $h \equiv h_i^{\;\;i}$. We raise and lower indices using the background metric $\delta_{ij}$.

The variable $\mu$ is fixed to \eqref{eq_mu_lambdas} as it is a constant. We keep $\nub$ in all calculations where the background otherwise vanishes and take the limit $\nub\rightarrow 0$ 
only at the end. We first calculate the secondary constraints, the first of which leads to $\Smu=0$ for the chosen function.
To expand the constraint \eqref{sec2}, that is $\Snu\approx 0$, to first order and determine $\neta$ we need $\Qcal$ and $ \Ucal $ to first order, however, since the latter is also needed for the Hamiltonian constraint
 to second order, we compute that to get
\begin{align}
\Qcal =&  \Qcal_0 \left(1 - \frac{1}{2} h  - \frac{\neta}{\nub}  + 8\pi\Gt  \frac{\varpi}{\Qcal_0\nub} \right),
\\
  \Ucal =&    \Qcal_0^2 \left[   \nub +   \neta  + \frac{\netat}{2\Kcal_2 } \right].
\end{align}
Thus, into \eqref{sec2} we find
\begin{align}
   \neta  = \frac{1}{ 1  + \frac{\nub}{2\Kcal_{2}  }} \left( 8\pi\Gt  \frac{\varpi}{\Qcal_0} - \frac{1}{2} \nub h   \right).
\end{align}

We now compute the diffeomorphism constraint \eqref{Hcal_i} which leads to
\begin{align}
N^i \Hcal_i \approx&  \frac{\Qcal_0 \nub}{8\pi \Gt}  h^i \grad_i\varphi - 2  h^i \grad_j \varpi^j_{\;\;i} 
\end{align}

Finally, we compute the Hamiltonian constraint \eqref{Hcal}. Since $\vec{A} \cdot \vec{D}\phi$ is second order, we need to zeroth order $C_2=(2-\KB)(2 + \KB \lambdas)/\KB$ 
and $C_1$ and $\Xi$ to second order. These are calculated as
\begin{align}
C_1 =& \frac{\Qcal_0}{8\pi\Gt} \nub + \varpi - \frac{2-\KB}{\KB}  \alpha^i\varpi_i 
\\
\Xi =&  \nub + \neta + \left[ \nub - \frac{(2-\KB)(2+\KB\lambdas)}{\KB} \right]  \alpha_i \alpha^i   
\end{align}
and into  \eqref{Hcal} we find
\begin{widetext}
\begin{align}
N\Hcal =&
  \frac{\Qcal_0^2}{8 \pi\Gt} \nub 
+  \Qcal_0 \varpi
+\frac{  \Qcal_0^2}{8\pi\Gt}   \nub  \Psi 
+ 8\pi \Gt \left[  2\varpi^{ij} \varpi_{ij}  -  \hat{\varpi}^2 + \frac{1}{2\KB}  |\vec{\varpi}|^2 
+  \frac{1}{4\Kcal_2}   \frac{1}{ 1  + \frac{\nub}{2\Kcal_{2}  }}   \varpi^2
\right]
- \frac{2-\KB}{\KB} \Qcal_0   \alpha^j\varpi_j 
\nonumber
\\
&
- \frac{ \Qcal_0 \nub}{4\Kcal_{2}  }  \frac{1}{ 1  + \frac{\nub}{2\Kcal_{2}  }}    \varpi h
- \frac{ \Qcal_0^2 \nub }{16\pi \Gt} \left[  
   \frac{1}{8}   \frac{1  - \frac{\nub}{2\Kcal_{2}  }  }{ 1  + \frac{\nub}{2\Kcal_{2}  }}   h^2
+ \frac{1}{4}  h^{ij}  h_{ij} 
+  |\vec{\alpha}|^2  
\right] 
+  \Qcal_0 \Psi\varpi
- \frac{\Qcal_0 \nub }{8\pi\Gt}    \vec{\alpha} \cdot \grad\phi
+  \Psi  \grad_i \varpi^i
-  \frac{2-\KB}{\KB}  \vec{\varpi} \cdot    \grad\varphi 
\nonumber
\\
&
+ \frac{1}{16\pi \Gt}
 \bigg[
 - \frac{1}{4} |\grad h|^2
+ \frac{1}{2}      \grad_i h \grad_j h^{ij} 
 - \frac{1}{4} \grad_k h^{ij} \left( \grad_i h^k_{\;\;j} + \grad_j h^k_{\;\;i} -  \grad^k h_{ij} \right)
+  \grad_i \Psi \grad_j h^{ij} - \grad \Psi \cdot \grad h 
\bigg]
\nonumber
\\
&
+ \frac{1}{16\pi \Gt}  \bigg[
 \KB  |\vec{B}|^2 
+ (2-\KB) \frac{2+ \lambdas\KB }{\KB}   |\grad\varphi + \Qcal_0 \vec{\alpha}|^2 
\bigg].
\end{align}
We may now set $\nub=0$ and combine the two to  get the second order Hamiltonian as
\begin{align}
 H^{(2)} =&
  \int d^3x\bigg\{  
 8\pi \Gt \left[  2\varpi^{ij} \varpi_{ij}  -  \hat{\varpi}^2 + \frac{1}{2\KB}  |\vec{\varpi}|^2 +  \frac{1}{4\Kcal_2}  \varpi^2 \right]
- \frac{2-\KB}{\KB}  \vec{\varpi}  \cdot \left(  \grad\varphi + \Qcal_0   \vec{\alpha} \right)
+ \frac{1}{16\pi \Gt}
 \bigg[
 - \frac{1}{4} |\grad h|^2
\nonumber
\\
&
+ \frac{1}{2}      \grad_i h \grad_j h^{ij} 
 - \frac{1}{4} \grad_k h^{ij} \left( \grad_i h^k_{\;\;j} + \grad_j h^k_{\;\;i} -  \grad^k h_{ij} \right)
+  \KB  |\vec{B}|^2 + (2-\KB) \frac{2+ \lambdas\KB }{\KB}   |\grad\varphi + \Qcal_0 \vec{\alpha}|^2 
\bigg]
\nonumber
\\
&
+  \Psi \left[\Qcal_0 \varpi +    \grad \cdot \vec{\varpi} + \frac{1}{16\pi \Gt} \left(   \grad^2 h - \grad_i  \grad_j h^{ij}  \right) \right]
 - 2  h^i \grad_j \varpi^j_{\;\;i} 
\bigg\}
\label{H_second}
\end{align}
\end{widetext}
Having found the second order Hamiltonian, we compare with the results from~\cite{SkordisZlosnik2021}, which only find the scalar mode Hamiltonian.
The Lagrangian and corresponding Hamiltonian in~\cite{SkordisZlosnik2021} is multiplied by $16\pi \Gt$ compared with what we have in this work, and we should do the same here
for proper comparison. Doing that leads to a rescaling of canonical momenta by  $16\pi \Gt$, that is, we define $\varpit^{ij} \equiv  16\pi \Gt \varpi^{ij}$ and similarly for $\varpi^i$ and $\varpi$.

We define the traceless operator $D_{ij} \equiv \grad_i \grad_j - \frac{1}{3} \grad^2 \delta_{ij}$ and 
expand the perturbations in scalar modes as
\begin{align}
h_{ij} =& - 2 \Phi \delta_{ij} + D_{ij} \eta
\\
\varpit^{ij} =& - \frac{1}{6}  P_\Phi \delta^{ij} + \frac{3}{2\grad^4}  D^{ij} P_{\eta} 
\\
A_i =& \grad_i \alpha
\\
\varpit^i =& - \grad^i \frac{1}{\grad^2} P_\alpha
\\
h_i =& -\grad_i \zeta
\end{align}
so that $\{\Phi,P_\Phi\}$,  $\{\eta,P_\eta\}$,  $\{\alpha,P_\alpha\}$ and  $\{\varphi,P_\varphi\}$ form canonical pairs, that is,
 $\int dt d^3x \left[\dot{h}_{ij} \varpi^{ij} + \dot{\alpha}_i \varpi^i \right]  = \int dt d^3x \left[\dot{\Phi}   P_\Phi  + \dot{\eta} P_{\eta} + \dot{\alpha}  P_\alpha\right]$.
Using the above expressions into \eqref{H_second} we find the scalar-mode Hamiltonian as
\begin{align}
 H^{(2)} =&
  \int d^3x\bigg\{  
- \frac{1}{24}  P_\Phi^2
+ \frac{3}{2}  \left| \frac{1}{\grad^2}  P_{\eta}\right|^2
+ \frac{1}{4\KB}  |\grad \frac{1}{\grad^2} P_\alpha  |^2 
\nonumber
\\
&
-  \frac{2-\KB}{\KB}    P_\alpha  \left(  \varphi + \Qcal_0 \alpha \right)
-2 \left| \grad  \left(\Phi + \frac{1}{6} \grad^2\eta \right) \right|^2 
\nonumber
\\
&
+  \frac{1}{8\Kcal_2}  P_\varphi^2 
 + (2-\KB) \frac{2+ \lambdas\KB }{\KB}   |\grad\left(\varphi + \Qcal_0 \alpha\right)|^2 
\nonumber
\\
&
+   \Psi \left[ \Qcal_0   P_\varphi  -  P_\alpha  - 4  \grad^2 \left(\Phi  + \frac{1}{6} \grad^2 \eta\right) \right]
\nonumber
\\
&
+ 2 \zeta \left(  \frac{1}{6} \grad^2  P_\Phi -  P_{\eta}  \right) 
\bigg\}
\end{align}
Our comparison is, however, not yet finished because in~\cite{SkordisZlosnik2021} one of the variables used is not $\varphi$ but the combination $\chi \equiv \varphi + \Qcal_0 \alpha$.
We thus perform a canonical transformation to new canonical pairs $\{\chi,P_\chi\}$ and $\{\alphat,P_\alphat\}$ defined through 
 $\alphat = \alpha$, $P_\chi = P_\varphi$  and $P_\alphat =  P_\alpha - \Qcal_0  P_\varphi$, and switch to Fourier space (where $\vec{k}$ denotes the Fourier wavevector) to get
\begin{widetext}
\begin{align}
 H^{(2)} =&
  \int \frac{d^3k}{(2\pi)^3}\bigg\{  
- \frac{1}{24}  \left|P_\Phi\right|^2
+ \frac{3}{2k^4}  \left|   P_{\eta}\right|^2
+  \frac{1}{8\Kcal_2}  \left|P_\chi\right|^2 
+ \frac{1}{4k^2\KB}  \left| \Qcal_0 P_\chi  +  P_\alphat   \right|^2 
-  \frac{2-\KB}{2\KB}   [  \chi \left(P_\alphat^* + \Qcal_0 P_\chi^*\right)  +  c.c. ]
\nonumber
\\
&
-2 k^2 \left| \Phi - \frac{1}{6} k^2\eta \right|^2 
 + (2-\KB) \frac{2+ \lambdas\KB }{\KB}  k^2  |\chi|^2 
+   \Psi C_\Psi^*
+   \Psi^* C_\Psi
+ \zeta C_\zeta^*
+ \zeta^* C_\zeta
\bigg\}
\label{H_final}
\end{align}
\end{widetext}
where
\begin{align}
C_\Psi \equiv& -\frac{1}{2} P_\alphat  + 2  k^2 \left(\Phi  - \frac{1}{6} k^2 \eta\right)  \approx 0
\\
C_\zeta \equiv& - \frac{1}{6} k^2  P_\Phi -  P_{\eta}  \approx 0
\end{align}
are two constraints imposed by the Lagrange multipliers $\Psi$ and $\zeta$, and where for brevity we use the same variable symbols in Fourier space as in real space.
The resulting Hamiltonian \eqref{H_final} is identical to the one found in~\cite{SkordisZlosnik2021}, up to some symbol relabeling.

\section{Conclusions}
\label{Conclusions}
In this article, we presented the general Hamiltonian analysis for AeST theory, which extends GR with the inclusion of a unit-timelike vector field $\Ah^\mu$ and scalar field $\phi$ in addition 
to the metric $g_{\mu\nu}$. To simplify the computations we further introduced two auxiliary fields $\mu$ and $\nu$ in order to avoid inversion of the free function $\Fcal$ which is part of the theory.
 Our analysis revealed the existence of 
four first class constraints and four second class constraints.
 The first class constraints consist of the three (primary) constraints $\Hcal_i$ defined in \eqref{Hcal_i} 
and the first class Hamiltonian $\Htot$ defined in \eqref{Htot},  
 which is the linear combination of the primary Hamiltonian constraint $\Hcal$ defined in \eqref{Hcal} (by itself second class), 
 $\Hcal_i$ and also $\Pmu$ and $\Pnu$.
The four second class constraints are the two canonical momenta $\Pmu$ and $\Pnu$  of the auxiliary fields $\mu$ and $\nu$, see \eqref{Pmu} and \eqref{Pnu},  
and the two secondary constraints $\Smu$ and $\Snu$ defined through \eqref{sec1} and \eqref{sec2}. The existence of these 
 second class constraints arises from the presence of the auxiliary fields $\mu$ and $\nu$. See table \ref{tab_constraints} for a summary of these cosntraints.

As we discussed in Sec.\ref{The_Theory}, the present theory defined by \eqref{NT_A_action} stems from reducing the more general model of~\cite{SkordisZlosnik2019} on the basis of
 the exact phenomenological requirements presented in Sec.\ref{sec:introduction}, as well as simplicity. 
It is possible that the assumption of having FLRW evolution close to $\Lambda$CDM may be relaxed, or that MOND can emerge in a way different than TeVeS theory and such 
possibilities will lead to a different action than \eqref{NT_A_action}.
It is also possible to extend  \eqref{NT_A_action} with higher-derivative terms in the scalar as in Horndeski~\cite{Horndeski:1974wa,Deffayet:2009mn}
 and more general theories~\cite{Gleyzes:2014dya,Langlois:2015cwa} or the vector as in~\cite{deRham:2020yet} or~\cite{Heisenberg:2018acv}, leading again to a richer phenomenology.
Our formalism can then be used to study such generalized cases  and determine their canonical structure.

We may use the constraint analysis to count the number of physical degrees of freedom.
We have six variables in the spatial metric $q_{ij}$, three in $A_i$ and one for each of $\phi$, $\mu$ and $\nu$, that is $12$ in total. Counting in the canonical momenta doubles this to $24$. 
We subtract the four second class constraints and twice the number of first class constraints which remove the gauge redundant degrees of freedom, that is, 
we subtract $12$ degrees of freedom because of the constraints. We finally divide by two to find six physical degrees of freedom.
It is interesting to compare this number to the result found in the case of SVT theories, specifically the case of broken gauge-invariance 
 which also propagate six degrees of freedom \cite{Heisenberg:2018acv}. However, as discussed in Section \ref{The_Theory}, there is no obvious overlap of AeST with SVT 
and there is no a-priori reason why the matching number of degrees of freedom would represent similar field structure.
We also note, that the additional degrees of freedom propagating may lead to interesting features, such as, additional polarization modes for gravitational waves, or perhaps,
new couplings to matter (which are not part of AeST). These could lead to important observational consequences which could then be used to put constraints on the theory.

Taking linear perturbations around a Minkowski background we expanded the Hamiltonian to quadratic order and recovered the same results found in~\cite{SkordisZlosnik2021} using different methods.
In the process we showed that the number of perturbative degrees of freedom found in~\cite{SkordisZlosnik2021} matches the number found here using the full non-linear theory.
Our formalism may be used to compute the quadratic Hamiltonian of AeST theory on other backgrounds in order to determine whether those backgrounds are stable or not. Of particular interest 
are the case of de Sitter space and static spherically symmetric configurations which we leave for a future work.

\acknowledgements
TZ was supported by the the Czech Science Foundation (GA\v{C}R) grant No. 20-28525S until 30.09.2022.,
and from 01.11.2022. by the project No. 2021/43/P/ST2/02141 co-funded by the Polish National Science 
Centre and the European Union Framework Programme for Research and Innovation Horizon 2020 under the 
Marie Sk\l{}odowska-Curie grant agreement No. 945339. 
C.S. acknowledges support by the European Structural and Investment Funds and the Czech Ministry of Education, Youth and Sports (MSMT) (Project CoGraDS-CZ.02.1.01/0.0/0.0/15003/0000437)
and by the Royal Society Wolfson Visiting Fellowship ``Testing the properties of dark matter with new statistical tools and cosmological data''.   

\appendix

\section{Useful Results}
\label{UsefulResults}
In the ADM formulation the vector field aligned with the time direction is
\begin{align}
t^{\mu} = (1 , \vec{0} )
\end{align}
while the normal to the hypersurface is decomposed as
\begin{align}
n^{\mu} =& \frac{1}{N} (1 , - \vec{N})  \;,
&
n_\mu =& (-N, \vec{0})
\end{align}
and the projector to the hypersurface as
\begin{align}
q^0_{\phantom{0}0} =& q^0_{\phantom{0}i} = 0  
&
q^i_{\phantom{i}0} =& N^i 
&
 q^i_{\phantom{i}j}= \delta^i_{\phantom{i}j}
\end{align}
The metric then has components
\begin{align}
g_{00} =& -N^2 + |\vec{N}|^2 \;,
&
 g^{00} =& -\frac{1}{N^2} \;,
\\
g_{0i} =& N_i \;,
&
 g^{0i} =& \frac{1}{N^2}N^i \;, 
\\
  g_{ij} =& q_{ij}
&
g^{ij}  =& q^{ij} - \frac{N^i N^j}{N^2}
\end{align}
The Christoffel connection splits into:
\begin{align}
\Gamma^0_{00} =& \frac{1}{N} \left[ \dot{N}  +  N^i D_i N  +   N^i  N^j  K_{ij}  \right]
\\
\Gamma^0_{0i} =& \frac{1}{N} \left[ D_i N  + N^j K_{ji}  \right]
\\
\Gamma^0_{ij} =&  \frac{1}{N} K_{ij}
\\
\Gamma^i_{00} =& \dot{N}^i + N^j  D_j N^i +2N K^{ij} N_j + N D^i N \nn\\
& -\frac{N^i}{N} \left[\dot{N} +   N^k N^l K_{kl} +  N^j  D_j N \right]
\\
\Gamma^i_{0j} =& N K^i_{\phantom{i}j} - \frac{1}{N} N^i N^k K_{kj} +   D_j N^i -\frac{N^i}{N} D_j N
\\
\Gamma^k_{ij} =& \gamma^k_{ij} - \frac{N^k}{N} K_{ij}
\label{del_Christoffel_ADM}
\end{align}
The components of the extrinsic curvature $K_{\mu\nu}$ can then be calculated to be:
\begin{align}
K_{00} =& N^i  N^j K_{ij}  
\\
K_{0i} = K_{i0} =& N^j K_{ij} 
\\
K_{ij} = N \Gamma^0_{ij}  =& \frac{1}{2N} \left( \dot{q}_{ij} - D_i N_j- D_j N_i  \right) 
\end{align}
The spacetime-covariant derivative of $\hat{A}^{\mu}$ is given by:
\begin{align}
\nabla_{\nu}\Ah^{\mu} &= \partial_{\nu}\Ah^{\mu} + \Gamma^{\mu}_{\nu\sigma}\Ah^{\sigma}
\end{align}
and individual components are
\begin{align}
\nabla_{0}\Ah^{0} =& \frac{\dot{\chi}}{N} +\frac{1}{N} A^i  \left( D_i N  + N^j K_{ij}  \right)
\\
\nabla_{i}\Ah^{0} =&  \frac{D_i \chi}{N} + \frac{1}{N} K_{ik} A^k 
\\
\nabla_{0}\Ah^{i} =&   \dot{A}^i
+  \left[ N K^i_{\phantom{i}j} +  D_j  N^i - \frac{N^i}{N} \left( N^k K_{kj} +   D_j N  \right) \right]A^j
\nn\\
&
 -  \frac{\dot{\chi}}{N} N^i + \chi \left( K^{ij} N_j +  D^i N \right)
\\
\nabla_{i}\Ah^{j} =&
 D_i A^j 
+ \chi   K^j_{\phantom{i}i} 
- \frac{N^j}{N} \left(  D_i \chi +  K_{ik}   A^k \right) 
\end{align}
From the above relations we then construct $J^\mu$ as
\begin{align}
J^0 =& 
 \frac{\chi}{N^2}  \left[ \dot{\chi} + A^i   D_i N -   N^i  D_i \chi \right]
\nonumber
\\
&
+ \frac{A^i}{N} \left[ K_{ij} A^j  + D_i \chi \right]
\label{J_0}
\\
J^i =& 
 \frac{\chi}{N}  \bigg\{  \dot{A}^i
+  \left[ N K^i_{\phantom{i}j} +  D_j  N^i - \frac{N^i}{N}   D_j N 
\right]A^j
\nn
\\
&
 -  \frac{\dot{\chi}}{N} N^i + \chi  D^i N 
-   N^j  \left[  D_j A^i - \frac{N^i}{N}  D_j \chi \right]
\bigg\}
\nn
\\
&
+  A^j   \left[  D_j A^i + \chi   K^i_{\phantom{i}j} - \frac{N^i}{N} \left(  D_j \chi +  K_{jk}   A^k \right)\right]
\label{J_i}
\end{align}

\section{Coefficients for the variations of the smeared Hamiltonian constraint}
\label{AB_coeff}
In evaluating the  smeared Hamiltonian constraint it is useful to define the coefficients $\Acal_{(Q_I)}$, $\Acal_{(Q_I)}^i$, $\Bcal_{(P_I)} $ and $\Bcal_{(P_I)}^i$
where  $Q_I = \{ q_{ij}, A_i, \phi\}$ and $P_I = \{ \Pi^{ij}, \Pi^i, \Pi\}$. See \eqref{Ham_Q_I} and \eqref{Ham_P_I}. Using 
\eqref{Hcal} and the variables $\Xi$, $C_1$, $C_2$ defined through \eqref{Xi_def}, \eqref{C_1_def} and \eqref{C_2_def} respectively
the $\Acal$-coefficients are found to be
\begin{widetext}
\begin{align}
\Acal_{(q_{ij})} =&   
  \frac{8\pi \Gt}{\sqrt{q}}\left\{ 
  4 \Pi^{k(i} \Pi_k^{\;\;j)} - 2 \hat{\Pi} \Pi^{ij} + \frac{1}{2\KB}  \Pi^i \Pi^j - \frac{C_1^2 C_2}{2\Xi^2}   A^i A^j 
 - \frac{1}{2}\left[ \Pi^{kl}  \left(  2\Pi_{kl}  -  \hat{\Pi}  q_{kl} \right) + \frac{1}{2\KB}  |\vec{\Pi}|^2 + \frac{C_1^2}{2\Xi} \right]q^{ij} 
\right\}  
\nonumber
\\
&
+ \frac{1}{2 \chi} \left(
 \frac{2-\KB}{\KB}  \vec{\Pi} \cdot    \vec{D}\phi 
-  \frac{C_1 C_2}{\Xi} \vec{A} \cdot \vec{D}\phi
-   \vec{D} \cdot \vec{\Pi}
\right)  A^i A^j 
- \frac{\chi C_1 C_2}{\Xi} \left( \frac{C_2}{\Xi} \vec{A} \cdot \vec{D}\phi A^i A^j  +   A^{(i} D^{j)} \phi \right)
\nonumber
\\
&
+  \frac{\sqrt{q}}{32\pi \Gt}\bigg\{ 
-  \Rcal  + 2 \Lambda - \KB  |\vec{B}|^2 
+ \left( \frac{C_2^2\chi^2}{\Xi}  + 2-\KB + \mu-\nu \right) \left(\vec{A}\cdot \vec{D}\phi \right)^{2} 
+  2  (2-\KB) \vec{A}\times \vec{B}  \cdot \vec{D}\phi 
\nonumber
\\
&
+ \left( 2-\KB + \mu + \frac{(2-\KB)^2}{\KB}  \chi^2 \right) |\vec{D}\phi|^2 
+ \Ucal
\bigg\} q^{ij} 
+ \frac{\sqrt{q}}{16\pi \Gt}\bigg\{ 
 R^{ij} + \KB  B^i B^j  
 - \frac{(2-\KB)^2}{\KB}   |\vec{D}\phi|^2 A^i A^j
\nonumber
\\
&
- 2 \left( \frac{C_2^2\chi^2}{\Xi}  + 2-\KB + \mu-\nu \right) \vec{A}\cdot \vec{D}\phi  A^{(i} D^{j)} \phi
+  2  (2-\KB) B_l \left( A_k \epsilon^{lk(i}   D^{j)} \phi -  \epsilon^{lk(i}  A^{j)}  D_k \phi \right)
\nonumber
\\
&
-  \frac{C_2^2}{\Xi^2} \left(2 \frac{2-\KB}{\KB} + \mu \right) \left(\vec{A}\cdot \vec{D}\phi \right)^{2} A^i A^j 
- \left( 2-\KB + \mu + \frac{(2-\KB)^2}{\KB}  \chi^2 \right) D^i \phi D^j \phi
\bigg\},
\end{align}
\begin{align}
 \Acal_{(A_i)} =&   
 \frac{8\pi \Gt}{\sqrt{q}} \, \frac{C_1}{\Xi} \left[
\frac{C_1 \, C_2}{\Xi} A^i
-  \frac{2-\KB}{\KB} \Pi^i
\right]
- \frac{1}{\chi}  \left\{ \frac{2-\KB}{\KB}  \vec{\Pi} \cdot    \vec{D}\phi -  \frac{C_1C_2}{\Xi}   \vec{A} \cdot \vec{D}\phi -   D_j \Pi^j \right\} A^i
\nonumber
\\
&
- \frac{\chi C_2}{\Xi} \bigg\{
  \frac{2-\KB}{\KB}   \vec{A} \cdot \vec{D}\phi  \Pi^i
-  C_1\left(  D^i\phi +   \frac{2C_2}{\Xi}   \vec{A} \cdot \vec{D}\phi A^i \right)
\bigg\}
\nonumber
\\
&
+  \frac{\sqrt{q}}{8\pi \Gt} \bigg\{  
 \epsilon^{ijk} \left[ \KB D_j B_k +    (2-\KB) B_j  D_k\phi \right]
+ \left[ \frac{C_2^2 \chi^2}{\Xi}  + 2-\KB + \mu-\nu \right] \vec{A}\cdot \vec{D}\phi  D^i \phi
\nonumber
\\
&
+   \frac{C_2^2}{\Xi^2}  \left[  2 \frac{2-\KB}{\KB} + \mu   \right] \left(\vec{A}\cdot \vec{D}\phi \right)^{2}  A^i
+ \frac{(2-\KB)^2}{\KB}  |\vec{D}\phi|^2   A^i
+    (2-\KB)  D_j (A^j  D^i\phi - A^i D^j \phi)  
\bigg\},
\end{align}
\begin{align}
 \Acal^i_{(A_j)} =&    
 \frac{\sqrt{q}}{8\pi \Gt}\left[ -\KB \epsilon^{ijk} B_k +(2-\KB)\left(\vec{A}^i \vec{D}^j\phi -\vec{A}^j \vec{D}^i\phi  \right) \right],
\end{align}
\begin{align}
\Acal^i_{(\phi)} =&   
   \frac{2-\KB}{\KB}  \chi  \Pi^i
-  \frac{\chi}{\Xi} C_1 C_2 A^i 
-  \frac{\sqrt{q}}{8\pi \Gt}\bigg\{ 
  (2-\KB)  \epsilon^{ijk}\vec{A}_j \vec{B}_k  
+\left[ \frac{\chi^2}{\Xi} C_2^2 + 2-\KB + \mu-\nu \right] \left(\vec{A}\cdot \vec{D}\phi\right) A^i
\nonumber
\\
&
+ \left( 2-\KB + \mu + \frac{(2-\KB)^2}{\KB}  \chi^2 \right) D^i\phi 
\bigg\}
\end{align}
\end{widetext}
and
\begin{align}
 \Acal^i_{(q_{jk})} =& 0 
\\
 \Acal_{(\phi)} =&   D_i \Acal^i_{(\phi)}.
\end{align}
The $\Bcal$-coefficients are
\begin{align}
 \Bcal_{(\Pi^{ij})} =& \frac{16\pi \Gt  }{\sqrt{q}}  \left(  2 \Pi_{ij} - \hat{\Pi} q_{ij} \right)
\\
 \Bcal^i_{(\Pi^{jk})} =&  0
\end{align}
\begin{align}
 \Bcal_{(\Pi^{i})} =& -  D_i \chi  - \chi  \frac{2-\KB}{\KB} \bigg\{ D_i\phi 
+  \frac{1}{\Xi}   C_2 \, \vec{A} \cdot \vec{D}\phi \; A_i
\bigg\}
\nonumber
\\
&
+ \frac{8\pi \Gt}{\KB \sqrt{q}}\bigg\{
 \Pi_i 
- \frac{2-\KB}{\Xi} C_1 A_i
\bigg\}
\\
 \Bcal^i_{(\Pi^{j})} =&  -\chi  \delta^i_{\;\;j}
\\
\Bcal_{(\Pi)} =& \frac{1}{\Xi} \bigg\{ \frac{8\pi \Gt  }{\sqrt{q}} C_1 +  \chi C_2 \, \vec{A} \cdot \vec{D}\phi \bigg\}
\\
\Bcal^i_{(\Pi)} =&   0
\end{align}

\bibliography{references}

\end{document}